\documentclass[10pt]{article}
\usepackage[utf8]{inputenc}
\usepackage{amsmath}
\usepackage{amssymb}
\usepackage{mathrsfs}
\usepackage{authblk}
\usepackage{color}
\usepackage[unicode,breaklinks=true]{hyperref}
\allowdisplaybreaks[1]

\usepackage{caption}
\usepackage{multirow}
\usepackage{subcaption}
\usepackage{upgreek}
\usepackage{cite}
\usepackage{xcolor}
\usepackage{graphicx}
\usepackage
[
a4paper,
left=1.8cm,
right=1.8cm,
top=3cm,
bottom=4cm,
]{geometry}
 \usepackage{subcaption}
\usepackage{graphicx}
\usepackage{epsfig}
\usepackage{amsfonts}
\usepackage{graphicx}
\usepackage{graphicx, epsfig, amssymb} 
\author[1]{Carlos Herdeiro}
\author[1]{Eugen Radu}
\author[1]{Etevaldo dos Santos Costa Filho}

\affil[1]{Departamento de Matemática da Universidade de Aveiro and
		Centre for Research and Development in Mathematics and Applications (CIDMA),
		Campus de Santiago, 3810-183 Aveiro, Portugal}

\begin{document}
\title{\bf Proca-Higgs balls and stars \\ in a UV completion for Proca self-interactions} 

\date{January 2023}

\maketitle

\begin{abstract}

We consider a Proca-Higgs model wherein a complex vector field gains mass via spontaneous symmetry breaking, by coupling to a real scalar field with a Higgs-type potential.
This vector version of the scalar Friedberg-Lee-Sirlin model, can be considered as a UV completion of a complex Proca model with self-interactions. We study the flat spacetime and self-gravitating solitons of the model, that we dub Proca-Higgs \textit{balls} and \textit{stars} respectively, exploring the domain of solutions and describing some of their mathematical and physical properties. The stars reduce to the well-known (mini-)Proca stars in some limits. The full model evades the hyperbolicity problems of the self-interacting Proca models, offering novel possibilities for dynamical studies beyond mini-Proca stars. 
\end{abstract}

\newpage

\tableofcontents

\newpage 

\section{Introduction}
Scalar boson stars (see~\cite{Jetzer:1991jr,Schunck:2003kk,Liebling:2012fv,Shnir:2022lba} for reviews) are a popular model of self-gravitating solitons that have found a variety of applications in gravitational physics, for instance in the context of dark matter, $e.g.$~\cite{Lee:1995af,Suarez:2013iw,Eby:2015hsq,Chen:2020cef}, and as black hole mimickers, $e.g.$~\cite{Schunck:1998cdq,Mielke:2000mh,Yuan:2004sv,Guzman:2005bs,Berti:2006qt,Guzman:2009zz,Vincent:2015xta,Sennett:2017etc,Grould:2017rzz,Olivares:2018abq,Herdeiro:2021lwl,Rosa:2022tfv}. The simplest and historically pioneering model is that of \textit{mini-boson stars}~\cite{Kaup:1968zz,Ruffini:1969qy}, which emerge for a massive, free, complex scalar field minimally coupled to Einstein's gravity. Adding self-interactions (or non-minimal couplings) allows a whole landscape of models with different properties and applications. In most models both static and spinning stars have been constructed - see $e.g.$~\cite{Colpi:1986ye,Ryan:1996nk,Schunck:1996he,Yoshida:1997qf,Schunck:2003kk,Kleihaus:2005me,Kleihaus:2007vk,Bernal:2009zy,Hartmann:2013tca,Grandclement:2014msa,Herdeiro:2016gxs,Herdeiro:2017fhv,Collodel:2017biu,Herdeiro:2018wvd,Alcubierre:2018ahf,Brihaye:2018grv,Herdeiro:2019mbz,Guerra:2019srj,Li:2019mlk,Kunz:2019sgn,Delgado:2020udb,Li:2020ffy,Zeng:2021oez,Brihaye:2021mqk,Herdeiro:2021jgc,Boskovic:2021nfs,Maso-Ferrando:2021ngp}. In some cases, multi-centre stars, which can be interpreted as several boson stars in equilibrium have also been reported~\cite{Herdeiro:2020kvf,Herdeiro:2021mol,Delgado:2021jxd,Gervalle:2022fze,Sun:2022duv,Cunha:2022tvk}. A key aspect of boson stars is that they are dynamically robust, for some models and in parts of the parameter space~\cite{Gleiser:1988rq,Lee:1988av,Seidel:1990jh,Kusmartsev:1990cr,Yoshida:1994xi,Guzman:2009xre,Sanchis-Gual:2019ljs,Kleihaus:2011sx,Siemonsen:2020hcg,DiGiovanni:2020ror,Sanchis-Gual:2021edp,Sanchis-Gual:2021phr}, allowing dynamical strong gravity studies, such as collisions and mergers of individual stars, to test their stability and obtain to gravitational wave templates - see $e.g.$~\cite{Balakrishna:2006ru,Palenzuela:2006wp,Bezares:2017mzk,Palenzuela:2017kcg,Croon:2018ftb,Bezares:2018qwa,Pacilio:2020jza,Helfer:2021brt,Jaramillo:2022zwg,Evstafyeva:2022bpr}. Scalar boson stars have flat spacetime counterparts if appropriate scalar self-interactions are included in the model; the corresponding flat spacetime solitons go by the name of \textit{$Q$-balls}~\cite{Coleman:1985ki}.

From a high energy physics viewpoint, most of the aforementioned models accommodating boson stars are regarded as effective field theories (EFTs) rather than fundamental ones, that require beyond the standard model constructions for a fundamental physics embedding. For instance, many of the self-interactions considered are non-renormalizable. Still, such EFTs are generically self-consistent, allowing not only the study of equilibrium solutions but also their dynamics, with or without gravity. For attempts to embed some of these EFTs in extensions of the standard model of particle physics see~\cite{Freitas:2021cfi}. 

Vector boson stars, on the other hand, have only been constructed more recently~\cite{Brito:2015pxa}. In analogy to their scalar cousins, they were initially constructed for a massive, free, complex field minimally coupled to Einstein's gravity. Such \textit{mini-Proca stars} share many of the properties of their scalar cousins - see $e.g.$~\cite{Sanchis-Gual:2017bhw,Cunha:2017wao,Herdeiro:2017fhv,DiGiovanni:2018bvo,Sanchis-Gual:2018oui,Herdeiro:2019mbz}. Some differences, however, can also be found, for instance in the geodesic structure of the spherical stars~\cite{Herdeiro:2021lwl} and in the stability properties of the spinning stars~\cite{Sanchis-Gual:2019ljs}. The latter has motivated phenomenological studies of collisions of spinning Proca stars and the comparison of the corresponding gravitational wave signals with some observed transients, in particular GW190521~\cite{CalderonBustillo:2020fyi,CalderonBustillo:2022cja}. In fact, a catalogue of gravitational waves from Proca stars has been reported~\cite{Sanchis-Gual:2022mkk}. As for their scalar cousins, under appropriate self-interactions \textit{Proca-balls} have been constructed in flat spacetime~\cite{Loginov:2015rya,Brihaye:2017inn,Heeck:2021bce,Dzhunushaliev:2021tlm}.

Once again, from the viewpoint of high energy physics, the simplest Proca model should be regarded as an EFT. The lack of gauge invariance, explicitly broken by the mass term, causes issues, such as non-renormalizability. In a more fundamental theory, the mass term should be obtained from a Higgs mechanism, as in the electroweak sector of the standard model. Still, as a classical EFT, the free Proca model is self-consistent. 

A natural next step considered Proca stars in models with self-interactions. These were reported in~\cite{Minamitsuji:2018kof,Herdeiro:2020jzx} for  quartic self-interactions. It has recently been pointed out, however, that generic self-interacting Proca fields (not necessarily complex, and even in flat spacetime) can suffer from a breakdown of hyperbolicity~\cite{PhysRevLett.129.151102,PhysRevLett.129.151103,Mou:2022hqb}. In other words, these models are not fully self consistent \textit{even as EFTs}. Since the theory is being taken beyond its regime of validity,  progress requires a completion of the theory, typically at high energies, thus a \textit{UV completion}.

UV completions of EFTs can be challenging. A notoriously difficult example is the quest for the UV completion of General Relativity, which has been a holy grail of theoretical physics for over half a century. But there are simpler field theories which breakdown as EFTs and have known possible UV completions. An example is a field theory with non-linear kinetic terms, used for $k$-essance~\cite{Lara:2021piy}. A similar construction to this case has been suggested, in fact,  for self-interacting Proca models~\cite{Barausse:2022rvg,PhysRevD.106.084022}. 

In this paper we shall consider a Proca-Higgs model that can both yield a mass to the Proca field via a Higgs mechanism -- thus addressing one of the obvious shortcomings of the Proca EFT -- and be considered as a UV completion of a self-interacting Proca model. We shall construct the balls ($i.e.$ flat spacetime solitons) and stars ($i.e.$ self-gravitating solitons) of this model, hereafter dubbed \textit{Proca-Higgs balls} and \textit{Proca-Higgs stars}, studying their physical and mathematical properties. As we shall see, the model approaches the free Proca model yielding mini-Proca stars in some limits; but unlike  the latter, it also contains flat spacetime solitons, by virtue of the (effective) self-interactions. Moreover, we show that the model is free of the hyperbolicity problems that plague the self-interacting Proca models, thus making it an interesting arena for exploring the dynamics of such Proca-Higgs solitons in a self-consistent manner.

This paper is organized as follows. In Section~\ref{sec2} we present the Proca-Higgs model that will be the focus of our work, its motivation and features. In Sections~\ref{sec3} and ~\ref{sec4} we discuss Proca-Higgs balls and stars, respectively. In Section~\ref{sec5} we discuss the compactness of the Proca-Higgs stars and some special circular geodesics, that could have impact in the stars'  phenomenology. In Section~\ref{sec6} we establish a no-hair theorem for spherical black holes with Proca-Higgs hair. We wrap up our results providing some discussion in Section~\ref{sec7}. In an Appendix we establish a first law type relation for the Proca-Higgs stars.

\section{The Proca-Higgs model}
\label{sec2}

\subsection{Action, field equations and limits}

We consider a model with a real scalar, $\phi$, and a complex vector field, $\mathcal{A}_\alpha$, both minimally coupled to Einstein's gravity. Unlike previous works dealing with Proca stars\footnote{These works consider the following action, that we shall refer to as the standard (complex) Proca model (where $\mu$ is a constant mass term)
\begin{equation}
\nonumber
\mathcal{S}=\int d^4x \sqrt{-g}\left[
\frac{R}{16 \pi  G}
-\frac{1}{4}\mathcal{F}_{\alpha\beta}\bar{\mathcal{F}}^{\alpha\beta}
-\frac{\mu^2}{2}\mathcal{A}_\alpha\bar{\mathcal{A}}^\alpha
\right]~. 
\end{equation}
} -- $e.g.$~\cite{Brito:2015pxa,Herdeiro:2016tmi} --, in this model the mass of the vector field results from a scalar-vector coupling. A non-zero vector mass results from the Higgs-like scalar potential, dynamically imposing a non-trivial scalar vacuum expectation value ($vev$) at infinity, $v=$constant.
Explicitly, the model is  described by the action:
\begin{equation}
\label{action}
\mathcal{S}=\int d^4x \sqrt{-g}\left[
\frac{R}{16 \pi  G}
-\frac{1}{4}\mathcal{F}_{\alpha\beta}\bar{\mathcal{F}}^{\alpha\beta}
-\frac{1}{2}\phi^2\mathcal{A}_\alpha\bar{\mathcal{A}}^\alpha
-\frac{1}{2} \partial_\alpha \phi \partial^\alpha \phi
-U(\phi)
\right]~, 
\end{equation}
where $R$ is the Ricci scalar of the spacetime metric $g_{\alpha\beta}$, with determinant $g$, and the vector field strength is $\mathcal{F}_{\alpha\beta}=\nabla_{\alpha}\mathcal{A}_{\beta}-\nabla_{\beta}\mathcal{A}_{\alpha}$, with overbar denoting complex conjugation. We shall focus on the `Mexican-hat' potential for the scalar field
\begin{equation}
~U(\phi)=\frac{\lambda}{4} (\phi^2-v^2)^2 \ ,
\end{equation}
where $\lambda>0$ is a constant.

The vector and scalar equations of the model are, respectively,
\begin{eqnarray}
&&
\label{procafe}
\nabla_\alpha\mathcal{F}^{\alpha\beta}=\phi^2\mathcal{A}^\beta \ ,
\\
&&
\label{scalarfe}
\Box\phi
=\frac{dU}{d\phi}
+\phi ~\mathcal{A}_\alpha\bar{\mathcal{A}}^\alpha,
\end{eqnarray}
whereas the Einstein equations read
\begin{equation}
R_{\alpha \beta}-\frac{1}{2}Rg_{\alpha \beta}=8 \pi G 
\left [
T_{\alpha \beta}^{(v)}
+
T_{\alpha \beta}^{(s)}
\right] \ ,
\label{Einstein-eqs}
\end{equation}
where the vector and scalar components of the energy-momentum tensor are
\begin{eqnarray}
&&
T_{\alpha\beta}^{(v)}=\frac{1}{2}
( \mathcal{F}_{\alpha \sigma }\bar{\mathcal{F}}_{\beta \gamma}
+\bar{\mathcal{F}}_{\alpha \sigma } \mathcal{F}_{\beta \gamma}
)g^{\sigma \gamma}
-\frac{1}{4}g_{\alpha\beta}\mathcal{F}_{\sigma\tau}\bar{\mathcal{F}}^{\sigma\tau}
+\phi^2\left[ \frac{1}{2}
(
\mathcal{A}_{\alpha}\bar{\mathcal{A}}_{\beta}
+\bar{\mathcal{A}}_{\alpha}\mathcal{A}_{\beta}
)
-\frac{1}{2}g_{\alpha\beta}\mathcal{A}_\sigma\bar{\mathcal{A}}^\sigma \right] \ ,
\\
&&
T_{\alpha\beta}^{(s)}=\partial_\alpha \phi \partial_\beta \phi-g_{\alpha \beta}
\left[\frac{1}{2} (\partial \phi)^2 +U(\phi) \right]~.
\end{eqnarray}
Here we have chosen to assign the mixed terms ($i.e.$ scalar-vector) to the vector energy-momentum tensor as to make it more Proca-like. This split, however, is arbitrary.

 The Proca-like vector field equations  (\ref{procafe})	imply the Lorenz-like condition, which is not a gauge choice, but rather a dynamical requirement:
	\begin{equation}
	\nabla_\alpha (\phi^2 \mathcal{A}^\alpha)= 0 \ .
	\label{lorentz}
	\end{equation}

Inspection of the field equations~\eqref{procafe}-\eqref{Einstein-eqs} reveals that $\phi=0$ is a consistent truncation of the model, yielding a complex (massless) vector minimally coupled to Einstein's gravity. In other words an Einstein-(double)Maxwell theory. On the other hand, $\phi=v\neq 0$ is \textit{not} a consistent truncation. The Einstein-vector equations yield an Einstein-complex-Proca system, but the scalar equation yields an additional  non-trivial constraint, $\mathcal{A}_\alpha\bar{\mathcal{A}}^\alpha=0$.

 Despite not reducing \textit{exactly} to the standard Proca model, there are limits in which the latter should be recovered. For asymptotically flat solutions, that are the focus of our study, the asymptotic behavior of the  scalar field is fixed by the condition
	\begin{equation}
	U(\phi) \to 0 \ , \qquad {i.e.}~~
	\phi  \to v~~{\rm as}~~r\to \infty \ ,
	\end{equation}
	which implies an effective mass $\mu\equiv v$ for the vector field. Thus, the standard Proca model~\cite{Brito:2015pxa,Herdeiro:2016tmi} is recovered asymptotically. One may also anticipate that the standard Proca model is recovered for ``large $\lambda$", as in this case, it becomes energetically costly for $\phi$ to depart from the vev. We shall see below how much this expectation is confirmed by the data.

On the other hand, the model~\eqref{action} will exhibit features unlike those of the standard (free) Proca model. Namely, flat-spacetime solitonic solutions (hereafter ``balls") may exist. This is suggested by the analogy with a (renormalizable) theory proposed by Friedberg, Lee and Sirlin \cite{Friedberg:1976me}.
	This model contains two scalars only,  and can be seen as the spin-zero version of (\ref{action}),
	the complex vector field being replaced by a complex scalar field $\psi$, with the matter Lagrangian density
	\begin{equation}
	\mathcal{L}=-\frac{1}{2}(\partial_\alpha \psi)(\partial^\alpha \psi)^*
	-\frac{1}{2}\phi^2\psi \psi^*
	-\frac{1}{2} \partial_\alpha \phi \partial^\alpha \phi
	-U(\phi) \ .
	\end{equation}
	The complex scalar $\psi$ becomes massive due to the coupling with the real scalar field $\phi$, which has a finite vev generated via a symmetry-breaking	potential. 	Since the Friedberg-Lee-Sirlin possesses particle-like solutions in a flat space background, it is natural to expect a similar result for its vector generalization that we are proposing. Below we shall confirm this expectation.

	The model~\eqref{action} possesses a  global $U(1)$ invariance of the complex vector field,  under the transformation 
	$\mathcal{A}_\mu\rightarrow e^{i\chi}\mathcal{A}_\mu$, with $\chi$ constant. This
	implies the existence of a conserved 4-current, 
	\begin{equation}
	\label{current}
	j^\alpha=\frac{i}{2}\left[\bar{\mathcal{F}}^{\alpha \beta}\mathcal{A}_\beta-\mathcal{F}^{\alpha\beta}\bar{\mathcal{A}}_\beta\right] \ .
	\end{equation}
	From (\ref{procafe}) it follows that $\nabla_\alpha j^\alpha=0$. Consequently, there exists a Noether charge, $Q$, obtained by integrating the temporal component of the 4-current on a space-like slice $\Sigma$:
	\begin{equation}
	Q=\int_\Sigma d^3x \sqrt{-g} j^0 \ .
	\label{q}
	\end{equation}

\subsection{A UV completion of a self-interacting Proca model}
The model we are considering~\eqref{action} can be regarded as a UV completion of a self-interacting Proca model. To see this, we start by expanding $\phi$ around its vev,
\begin{align}
    \phi=v(1+\rho)\ ;
\end{align}
one gets
\begin{align}
    \frac{1}{2}\partial_\alpha\phi\partial^\alpha\phi+U(\phi)
    =v^2\left(\frac{1}{2}\partial_\alpha\rho\partial^\alpha\rho+\lambda v^2\rho^2+\lambda v^2\rho^3+\frac{\lambda v^2}{4}\rho^4\right)\ ,
\end{align}
which means that the perturbation around the vev, $\rho$, acquires an effective mass $M_\rho\equiv\sqrt{2\lambda}v$. The equation of motion for $\rho$ then reads
\begin{align}
    \Box\rho=\frac{M_\rho^2}{2}\rho(1+\rho)(2+\rho)+(1+\rho)\mathcal{A}_\alpha\Bar{\mathcal{A}}^\alpha \ .
    \label{eq:3}
\end{align}
At energies much lower than $M_\rho$, one can freeze the scalar degree of freedom ($\Box\rho\approx0$) and solve \eqref{eq:3} for $\rho$
\begin{align}
    \rho_\pm=-1\pm\sqrt{1-\frac{2\mathcal{A}_\alpha\Bar{\mathcal{A}}^\alpha}{M_\rho^2}} \ .
    \label{eq:2}
\end{align}
Using \eqref{eq:2} in the initial model \eqref{action}, one obtains the low energy EFT
\begin{align}
    \mathcal{S}_\text{eff}
    &=\int\text{d}^4x\left[\frac{R}{16\pi G}-\frac{1}{4}\mathcal{F}_{\alpha\beta}\Bar{\mathcal{F}}^{\alpha\beta}-\frac{v^2}{2}\left(\mathcal{A}_\alpha\Bar{\mathcal{A}}^\alpha-\frac{(\mathcal{A}_\alpha\Bar{\mathcal{A}}^\alpha)^2}{M_\rho^2}\right)\right]\ .
\end{align}
This is a complex Proca model with quartic self-interactions. We had mentioned before that setting $\phi=v$ is not a consistent truncation of the initial model~\eqref{action}. We have just shown that considering small oscillations around the vev, instead of getting a nuisance constraint from the scalar equation, that constraint can be used to get an extra interaction in the model, which may suffice to describe the dynamics at sufficiently low energies. It is worth noticing that the effective Proca field theory obtained has a definite sign for the self-interactions, $i.e.$ it reads
\begin{align}
    \mathcal{L}^{\rm Proca \ EFT}=-\frac{1}{4}\mathcal{F}_{\alpha\beta}{\mathcal{F}}^{\alpha\beta}-\frac{\mu^2}{2}\mathcal{A}_\alpha\Bar{\mathcal{A}}^\alpha-\alpha_2(\mathcal{A}_\alpha\Bar{\mathcal{A}}^\alpha)^2\ ,
    \label{Paction2}
\end{align}
with the mass term $\mu^2=v^2$, determined by the scalar vev, and the self interactions
\begin{equation}
\alpha_2=-\frac{v^2}{2M_\rho^2}=-\frac{1}{4\lambda}<0 \ ,
\end{equation}
determined by the strength of the scalar self-interactions. Again, we see that for $\lambda$ ``large", the free Proca model should be approximately recovered.

We remark that similar considerations have been made in  \cite{PhysRevD.106.084022}. Our concrete model is distinct from those considered in this reference and our goal includes acquiring mass dynamically.


\subsection{The hyperbolicity issue}


The dynamics of self-interacting vector fields can sometimes be written as if regulated by a so-called effective metric \cite{Coates:2022nif,PhysRevLett.129.151103,PhysRevLett.129.151102}, that depends on the vector field itself and on the spacetime, which is a fundamental aspect for understanding the problem. In other words, such an effective metric controls the principal part of the differential operator governing the vector field, and it can lead to a loss of hyperbolicity. Given the hyperbolicity issues observed for self-interacting Proca fields~\cite{Coates:2022nif,Mou:2022hqb,PhysRevLett.129.151103,Barausse:2022rvg,PhysRevLett.129.151102}, and given the connection between model~\eqref{action} and the Proca model with quartic self-interactions demonstrated in the last subsection, it is relevant to discuss the hyperbolicity of model~\eqref{action}.

In this subsection, we investigate hyperbolicity issues in the Proca-Higgs  model, applying the methodology introduced in \cite{Coates:2022nif,PhysRevLett.129.151103,PhysRevLett.129.151102}, to obtain the effective metric ruling the propagating of the vector field. This approach is focused on the principal part of the differential operator. It does not rule out tachyonic instabilities \cite{PhysRevLett.129.151102}, which are of a different nature and do not signal a breakdown of the EFT.

For this analysis we consider a real vector field $\mathcal{A}^\mu$ with norm $ \mathcal{A}^2= \mathcal{A}^\mu \mathcal{A}_\mu$, coupled with a real scalar field, $\phi$, described by the Lagrangian density
\begin{equation}
	\mathcal{L}= 
	\frac{R}{16 \pi  G}
	-\frac{1}{4}\mathcal{F}_{\alpha\beta}\mathcal{F}^{\alpha\beta}
	-\dfrac{1}{2}V(\phi,\mathcal{A}^2)
	-\frac{1}{2} \partial_\alpha \phi \partial^\alpha \phi
	-U(\phi) \ .
\end{equation}
We assume that $\phi$ only occurs multiplied by powers of $\mathcal{A}^2$ in the potential $V(\phi,\mathcal{A}^2)$. The vector field equation is then

\begin{equation}\label{VFE}
	- \nabla_{\mu}\nabla^{\mu}\mathcal{A}_{\nu} + \nabla_{\mu}\nabla_{\nu}\mathcal{A}^{\mu} + \mathcal{A}_{\nu} \
	V^{(0,1)}=0 \ ,
\end{equation}
where $ V^{(n,m)}\equiv  \dfrac{\partial^n\partial^m V\bigl(\phi ,  \mathcal{A}^2\bigr)}{(\partial\phi)^n \ (\partial\mathcal{A}^2)^m})$. In order to construct the effective metric, we want to write the second term in \eqref{VFE} as a wave operator. By taking the divergence of the above equation, one finds the modified Lorenz gauge condition
\begin{equation}
	\nabla^{\nu}\left(\mathcal{A}_{\nu}V^{(0,1)}\right)=0 \ ,
\end{equation}
which can be re-written as
\begin{equation}\label{expandgauge}
	\nabla_{\mu }\mathcal{A}^{\mu } V^{(0,1)} + 2 \mathcal{A}^{\mu } \mathcal{A}^{\nu } \nabla_{\mu }\mathcal{A}_{\nu } V^{(0,2)} + \mathcal{A}^{\mu } \nabla_{\mu }\varphi V^{(1,1)}=0 \ .
\end{equation}
Next, by using the definition of the Riemann curvature tensor $
	\left(\nabla_{\mu}\nabla_{\nu}-\nabla_{\nu}\nabla_{\mu}\right)\mathcal{A}^\mu=R_{\nu\mu}\mathcal{A}^{\mu}$
and using equation \eqref{expandgauge}, we can cast the vector field equation \eqref{VFE} as

\begin{multline}\label{wavegeneral}
	\nabla_{\mu }\nabla^{\mu }\mathcal{A}_{\nu }+\frac{2 \nabla_{\alpha }\nabla_{\nu }\mathcal{A}_{\mu } V^{(0,2)}}{V^{(0,1)}}A^{\mu}\mathcal{A}^{\alpha} -  \mathcal{A}_{\nu } V^{(0,1)} + \frac{\nabla_{\mu }\phi \nabla_{\nu }\mathcal{A}^{\mu } V^{(1,1)}}{V^{(0,1)}}- R_{\nu \mu }\mathcal{A}^{\mu}\\
	+\frac{1}{V^{(0,1)}}\left\{  2\left( \nabla_{\alpha }\mathcal{A}_{\mu } +  \nabla_{\mu }\mathcal{A}_{\alpha }\right) \nabla_{\nu }\mathcal{A}^{\alpha } V^{(0,2)}+ V^{(1,1)}\nabla_{\nu }\nabla_{\mu }\phi  +\left( V^{(2,1)}-  \frac{ \Bigl(V^{(1,1)}\Bigr)^2}{V^{(0,1)}}\right)\nabla_{\mu }\phi \nabla_{\nu }\phi\right\}\mathcal{A}^{\mu}\\
	+\frac{2}{V^{(0,1)}}\left\{ -  \frac{  V^{(0,2)} V^{(1,1)}}{V^{(0,1)}}(  \nabla_{\alpha }\phi \nabla_{\nu }\mathcal{A}_{\mu }  +  \nabla_{\alpha }\mathcal{A}_{\mu } \nabla_{\nu }\phi  )+ V^{(1,2)}(\nabla_{\alpha }\phi \nabla_{\nu }\mathcal{A}_{\mu }  + \nabla_{\alpha }\mathcal{A}_{\mu } \nabla_{\nu }\phi )\right\}\mathcal{A}^{\mu}\mathcal{A}^{\alpha}\\
	+ \frac{4 \nabla_{\beta }\mathcal{A}_{\alpha } \nabla_{\nu }\mathcal{A}_{\mu } }{V^{(0,1)}}\left\{ V^{(0,3)}- \frac{ \Bigl(V^{(0,2)}\Bigr)^2}{V^{(0,1)}} \right\}\mathcal{A}^{\mu}\mathcal{A}^{\alpha}\mathcal{A}^{\beta }=0 \ .
\end{multline}
We can now substitute the second term in \eqref{wavegeneral} by using $\mathcal{F}_{\alpha\beta}=\nabla_{\alpha}\mathcal{A}_{\beta}-\nabla_{\beta}\mathcal{A}_{\alpha}$ and, consequently, the principal part of the differential operator becomes 
\begin{equation}
	\nabla_{\mu }\nabla^{\mu }\mathcal{A}_{\nu }+2\frac{ \nabla_{\alpha }\nabla_{\nu }\mathcal{A}_{\mu } V^{(0,2)}}{V^{(0,1)}}A^{\mu}\mathcal{A}^{\alpha} = \hat{g}_{\mu\alpha}\nabla^{\mu }\nabla^{\alpha }\mathcal{A}_{\nu }+2\overbrace{\frac{ \nabla_{\alpha }\mathcal{F}_{\nu\mu} V^{(0,2)}}{V^{(0,1)}}A^{\mu}\mathcal{A}^{\alpha} }^{\Theta_\nu} .
\end{equation}
We have introduced the effective metric
\begin{equation}
	\hat{g}_{\mu\nu}\equiv g_{\mu\nu}+\frac{2 V^{(0,2)}}{V^{(0,1)}}\mathcal{A}_{\mu}\mathcal{A}_{\nu} \ .
\end{equation}

Similarly as \cite{PhysRevLett.129.151103}, the term $\Theta_\nu$ contributes to the principal part of the differential operator and must be taken into account. As in the pure Proca case, Eq. \eqref{wavegeneral} is not manifestly hyperbolic. Moreover, similar procedures to those in Appendix B of \cite{PhysRevLett.129.151103} shows that, in fact, the effective metric, $\hat{g}_{\mu\nu}$, governs the dynamics of the vector field and change of its signature leads to a breaking into the hyperbolicity of the equation.

Notice, however, that our model is free of hyperbolicity issues caused by change in signature of this effective metric, since for this particular model the effective metric reduces to the spacetime one. This is verified for our model~\eqref{action}, for which 
\begin{equation}
V(\phi,\mathcal{A}^2)=\phi^2\mathcal{A}^2\ \qquad \Rightarrow \qquad  V^{(0,2)}=0 \ ,
\end{equation}
which confirms the model~\eqref{action} is free of hyperbolicity issues. In this case, moreover,  \eqref{wavegeneral} simplifies to
\begin{equation}
	\nabla_{\mu }\nabla^{\mu }\mathcal{A}_{\nu } - \mathcal{A}^{\mu } R_{\nu \mu } -  \mathcal{A}_{\nu } \phi^2  + 2  \nabla_{\nu }\mathcal{A}^{\mu } \nabla_{\mu }\ln\phi+ 2 \mathcal{A}^{\mu } \nabla_{\nu }\nabla_{\mu }\ln\phi=0 \ .
\end{equation}
On the other hand, the effective mass matrix
\begin{equation}
	\mathcal{M}_{\mu\nu}= R_{\nu \mu } +  \delta_{\mu\nu } \phi^2  - 2  \nabla_{\nu }\nabla_{\mu }\ln\phi \ ,
\end{equation}
can have vanishing determinant. Hence the model can present tachyonic instabilities.

\subsection{Ansatz, units and numerical approach}   
Throughout this paper we shall use units with $c=1$. In the explicit computation of the solutions below, we shall focus on spherically symmetric solutions. Then,  by using standard coordinates $(t,r,\theta,\varphi)$, we shall take the following ansatz for the ``matter" fields
\begin{equation}
\mathcal{A}=e^{-i \omega t}\left[f(r)dt+i g(r)dr\right]\,,\qquad \phi=\phi(r)\,.
\label{matteransatz}
\end{equation}
Here, $f$, $g$ and $\phi$ are all real functions, which only depend on the radial coordinate $r$, and $\omega$ is a real frequency parameter, assumed to be non-negative without loss of generality.\footnote{The electromagnetic potential $g$ should not be confused with the determinant of the metric.}
When computing self-gravitating solitons (``stars") we  assume the following form for the line element in isotropic coordinates
\begin{equation}\label{metric}
ds^2=-e^{2 F_0(r)}dt^2+e^{2 F_1(r)}\left[dr^2+r^2( d\theta^2+ \sin ^2\theta d\varphi^2)\right] \ ,
\end{equation}
where $F_0$ and $F_1$ are radial functions.

We shall now discuss some convenient rescalings of the variables and parameters of the model, that shall be used in the remainder of the paper. Unlike the standard (free) Proca model, the theory~\eqref{action} allows for flat spacetime solitons. To simplify comparisons between such Proca-Higgs balls and the self-gravitating stars, we introduce the dimensionless quantities
\begin{equation}
r\rightarrow \dfrac{r}{v}\,, \qquad \omega\rightarrow v \omega \,,\qquad\phi\rightarrow v\phi\, , \qquad  \mathcal{A}_{\alpha}\rightarrow v \mathcal{A}_{\alpha}\,,
\end{equation}
such that the field equations~\eqref{procafe}-\eqref{Einstein-eqs} become
\begin{eqnarray}
&&
\nabla_\alpha\mathcal{F}^{\alpha\beta}=\phi^2\mathcal{A}^\beta \ ,
\label{v2}
\\
&&
\Box\phi
=\lambda(\phi^2-1)\phi
+\phi ~\mathcal{A}_\alpha\bar{\mathcal{A}}^\alpha \ ,
\label{s2}
\\
&&
R_{\alpha \beta}-\frac{1}{2}Rg_{\alpha \beta}=2\alpha^2
\left[
T_{\alpha \beta}^{(v)}
+
T_{\alpha \beta}^{(s)}
\right] \ ,
\label{e2}
\end{eqnarray}
where we have defined the coupling constant 
\begin{equation}
\alpha^2\equiv 4\pi G v^2 \ .
\end{equation}
This scaling reduces the number of tuneable  couplings/parameters in the action from 3 $(G,\lambda,v)$ to 2 $(\alpha,\lambda)$. Then, the (gravitational) decoupling limit -- in particular to be used for a flat spacetime background to compute balls -- amounts to $\alpha=0$. 

Under this scaling the Noether charge remains invariant
\begin{equation}
Q=\int_\Sigma d^3x \sqrt{-g} j^0 \ .
\label{qint}
\end{equation}
The mass/energy of balls and the Komar  mass of the stars, on the other hand, scales as $M\rightarrow Mv$. In the case of stars, $M$ can be computed in two ways: $(1)$ the Komar mass
\begin{equation}\label{mint}
M_{\text{Komar}}\equiv - \int_\Sigma d^3x \sqrt{-g} \left(2T_{0}^{\ 0}-T\right)\, ,
\end{equation}
where $T$ is the trace of the \textit{total}  energy-momentum tensor. Obviously, this is also applicable to balls, in which case it reduces to the standard integral over the energy density\footnote{This can be seen from~\eqref{mint} together with the Deser/virial identity for flat spacetime solitons, which establishes that the integral of the spatial trace on a spacelike slice vanishes~\cite{Herdeiro:2022ids}.}
\begin{equation}\label{mint2}
M_{\text{Komar}}^{\rm balls}\equiv  \int_\Sigma d^3x \sqrt{-g} \, T_{00}\, ;
\end{equation}
 and
$(2)$ the ADM mass, that can be read off from the metric behavior at infinity
\begin{equation}
-g_{tt} \underset{r\rightarrow\infty}{\approx} 1 - \dfrac{2 G M_\text{ADM}}{r} \ .
\end{equation}
Comparing $M_{\rm ADM}$ with $M_{\rm Komar}$, for stars, provides a test for the numerical quality of the solutions. 

Turning now to the issue of numerical construction of the solutions, we mention that our
approach is similar for both Proca-Higgs balls and stars.
The solutions are found by solving a set of  coupled non-linear ordinary differential equations 
for the functions  ${\cal F} =(F_0, F_1; \phi,f, g)$
(with $F_0=F_1=0$ for balls),
 which are displayed in Sections~\ref{eq-Asymptotic_flat} and \ref{sec4-eq}.
 These equations were subject to the boundary conditions introduced in the aforementioned Sections. 
 The professional package \textsc{fidisol/cadsol} \cite{SCHONAUER1989279},  which employs a finite difference method with an arbitrary grid and arbitrary consistency order, has been used to perform all numerical calculations reported in this work\footnote{We mention that some of the solutions have been recovered by using a Runge-Kutta ordinary differential equations solver and a shooting technique.}. This solver uses a Newton-Raphson method, which requires a good first guess in order to start a successful iteration procedure (see the references \cite{Herdeiro_2015,Delgado:2022pwo} for a more in-depth explanation of the solver).
Inside the solver, we introduce a
compactified radial variable $ x=r/(c+r)$
 (with $c$ an input parameter which is usually taken equals to one)
which maps the semi-infinite region $[0,\infty)$ to the finite region $[0,1]$, and make the substitutions 

$${\cal F}_{,r} \longrightarrow \frac{1}{c}(1- x)^2 {\cal F}_{, x},\qquad {\cal F}_{,r r} \longrightarrow \frac{1}{c^2} (1- x)^4  {\cal F}_{, x  x} - \frac{2}{c^2} (1- x)^3 {\cal F}_{, x }.$$ 

We then discretize the equations for ${\cal F}$ on a grid in $x$. The results in this work have been found for an equidistant grid varying from (around) 400 to 800 points, covering the integration region $0\leq  x \leq 1$. 

The solver also provides error estimates for each unknown function, which allows for judging the quality of the computed solution. The numerical error for the solutions reported in this work is estimated to be typically  $<10^{-5}$. Moreover, we also use physical constrain to test the trustworthiness of the numerical solutions, e.g., the equivalence between the ADM mass and the Komar one, the virial identity, the first-law identity and the gauge condition.

\section{Proca-Higgs balls}
\label{sec3}
\subsection{The equations and asymptotics}
\label{eq-Asymptotic_flat}

We shall now present our numerical results on the solitonic solutions of~\eqref{action}. First we consider the flat spacetime solutions in the decoupling limit: we set $\alpha=0=F_0=F_1$. Then,  the only field equations to solve are \eqref{v2} and \eqref{s2}, which, under the ansatz~\eqref{matteransatz}, yield
\begin{equation}\label{EQM1}
\frac{d}{d r}\left\{r^{2}\left[f^{\prime}-\omega g\right]\right\}=r^2 \phi^2f\,,\qquad \omega g-f'=\dfrac{\phi^2 g}{\omega} \ ,
\end{equation}
\begin{equation}\label{EQM2}
\phi''= - \left(f^2-g^2+\lambda \right)\phi-\frac{2 \phi'}{r}+\lambda  \phi^3 \ ,
\end{equation}
where the prime denotes the derivative with respect to $r$. These equations are constrained by the gauge condition \eqref{lorentz}, which becomes
\begin{equation}
g'+\omega f=-\frac{2 g\left(r \phi'+\phi\right)}{r \phi} \ .
\end{equation}

For the Proca-Higgs balls, the globally conserved quantities, Noether charge~\eqref{qint} and mass/energy~\eqref{mint2}, become
\begin{align}
&Q=4\pi\int_{0}^{\infty}\frac{ g^2\phi^2}{\omega} r^2dr\,,\\ &M=2\pi v\int_{0}^{\infty}\left[\left(f'-\omega g\right)^2+\phi^2 \left(f^2+g^2\right)+\frac{1}{2} \lambda \left(\phi^2-1\right)^2+ \phi'^2\right]r^2dr\, .
\end{align} 
These solutions obey the virial identity (see $e.g.$~\cite{Herdeiro:2021teo})
\begin{equation}
\int_{0}^{\infty} \left\{-\dfrac{1}{2}\phi'^2-\dfrac{3\lambda}{4}\left(\phi^2-1\right)^2+\dfrac{1}{2}\phi^2\left[3 f^2 + g^2\left(-1+\dfrac{\phi^2}{\omega^2}\right)\right]\right\}r^2dr=0\,,
\label{virialflat}
\end{equation}
which can be used to test the numerical accuracy of the solutions. All solutions reported in this work obey the virial identity up to errors of order $10^{-5}-10^{-8}$; the larger errors are for solutions occurring inside the spiral.

To numerically integrate eqs.~\eqref{EQM1}-\eqref{EQM2} 
we need to impose boundary conditions. At the origin, $r = 0$, these establish regularity and read,
\begin{align}
&\phi(r)=\phi_0-\frac{\phi_0}{6}  \left(f_0^2-\lambda( \phi_0^2-1)\right)r^2+\mathcal{O}(r^3)\,,\\
&f(r)=f_0+\frac{f_0}{6}  \left(\phi_0^2-\omega^2\right)r^2+\mathcal{O}(r^3)\,,\\
&g(r)=-\frac{f_0 \omega}{3}r+\mathcal{O}(r^3)\,.
\end{align}
As such, $\phi(r=0)=\phi_0$ and $f(r=0)=f_0$. Due to the $\mathbb{Z}_{2}$ symmetry for the scalar field and the global $U(1)$ symmetry for the vector potential, $\phi_0$ and $f_0$ can be chosen positive. On the other hand, we impose that the scalar field reaches its vev when $r\rightarrow\infty$ and the vector potentials decay exponentially, due to the effective asymptotic mass term:
\begin{align}
&\phi(r)= 1+\frac{c_1 e^{- \sqrt{2\lambda} r}}{r}+\cdots\,,\label{phiinf}\\
&f(r)=\frac{c_2 e^{-r \sqrt{1-\omega^2}}}{r}+\cdots\,,\label{finf}\\
&g(r)= -\frac{c_2 \omega e^{-r \sqrt{1-\omega^2}} \left(r \sqrt{1-\omega^2}+1\right)}{r^2 \left(\omega^2-1\right)}+\cdots\,,\label{ginf}
\end{align}
where $c_1$ and $c_2$ are arbitrary real constants. One observes that in order to have localised solutions, the frequency is bounded, $\omega<1$. This is the standard bound state condition, since our scaling made the asymptotic effective mass term in the vector equation equal to unity.

\subsection{Numerical Results}

\begin{figure}[h!]
     \includegraphics[width=0.47\textwidth]{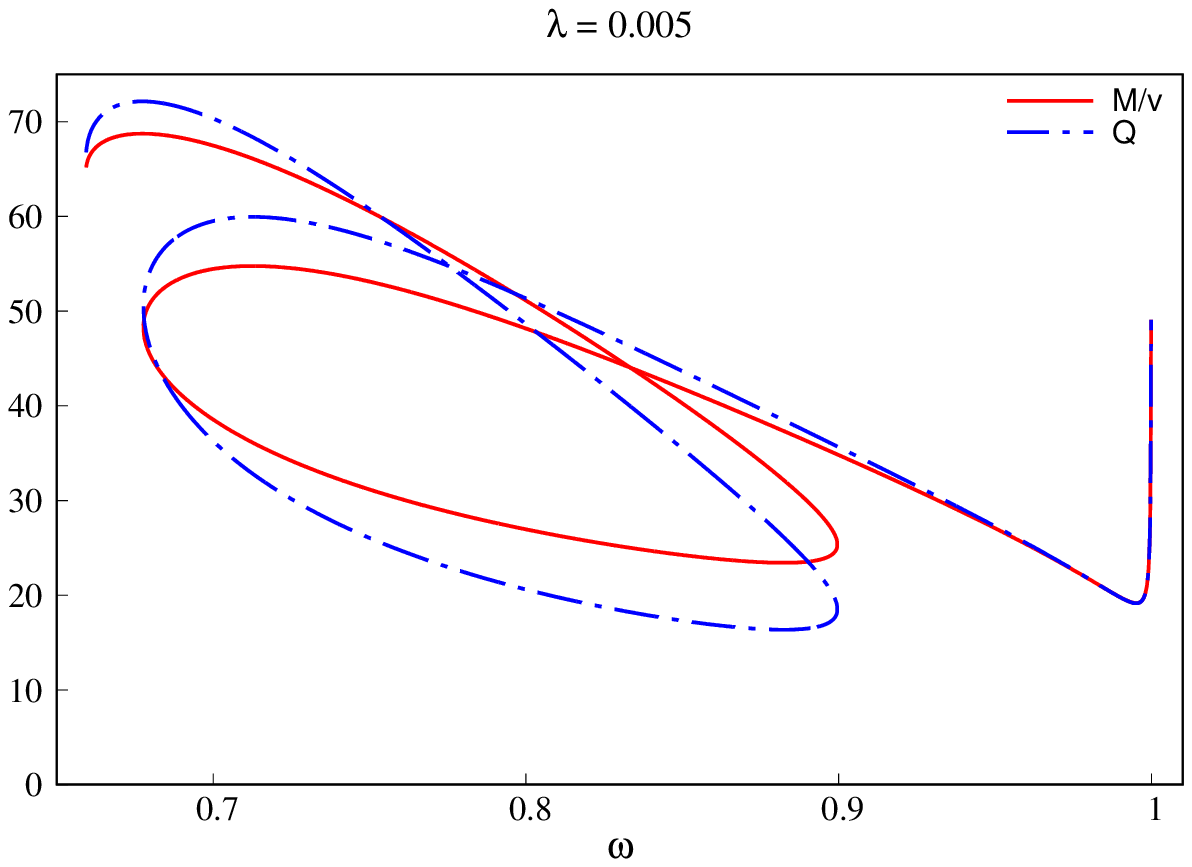} \ \ \ \ 
 \includegraphics[width=0.47\textwidth]{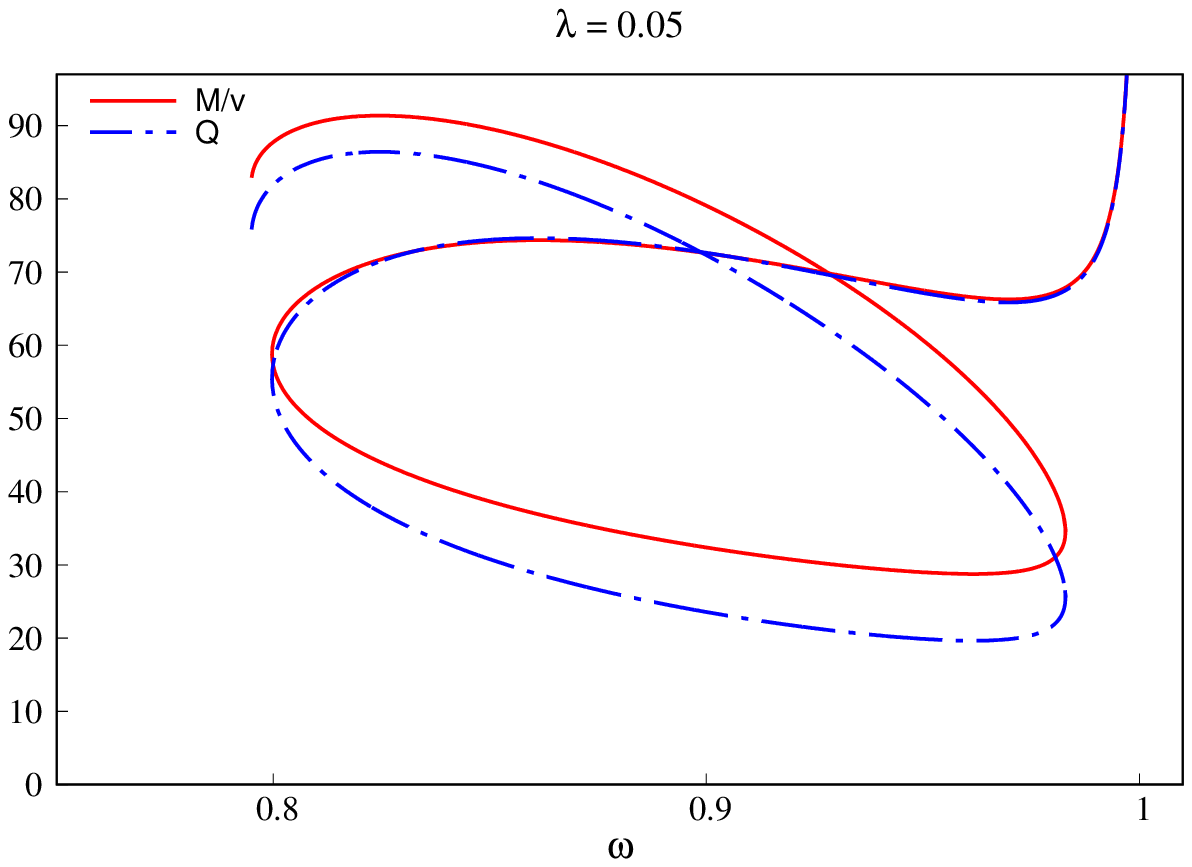}\\
	\includegraphics[width=0.48\textwidth]{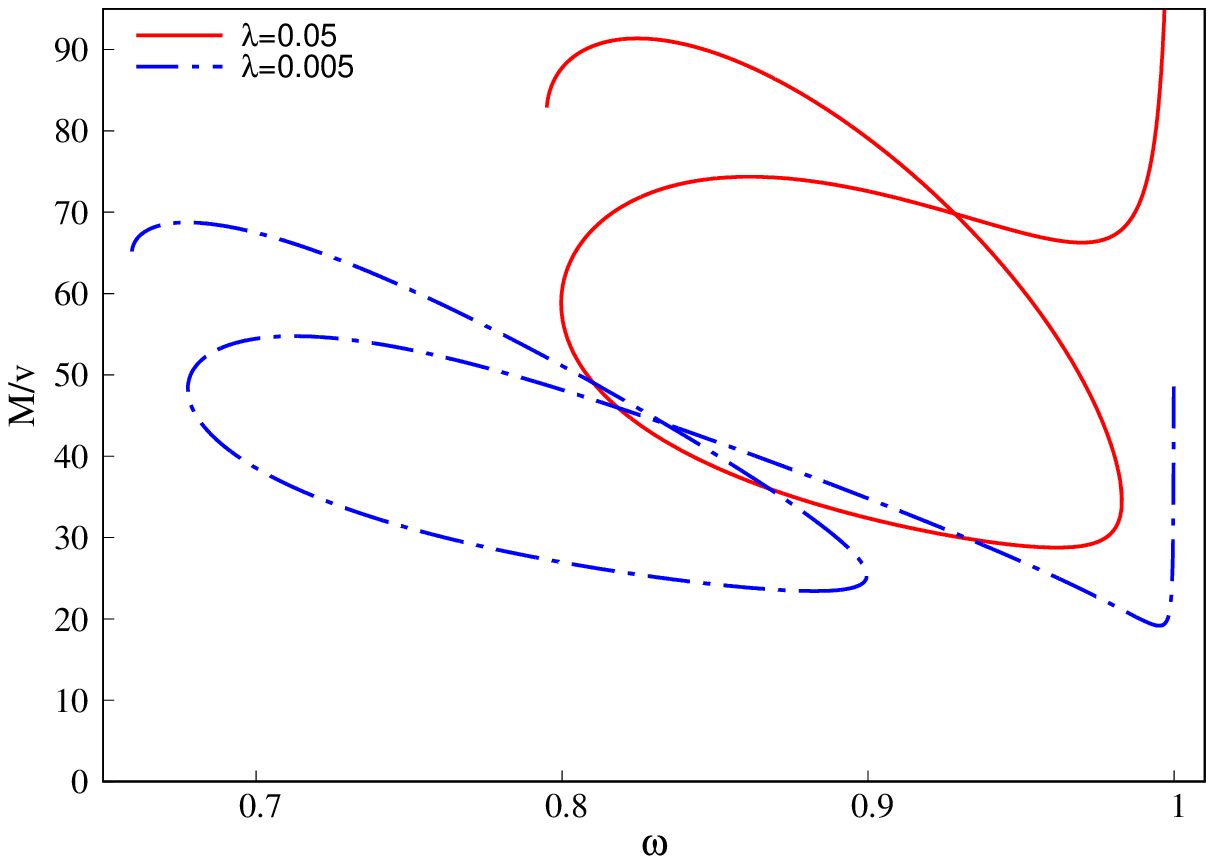}
 \includegraphics[width=0.485\textwidth]{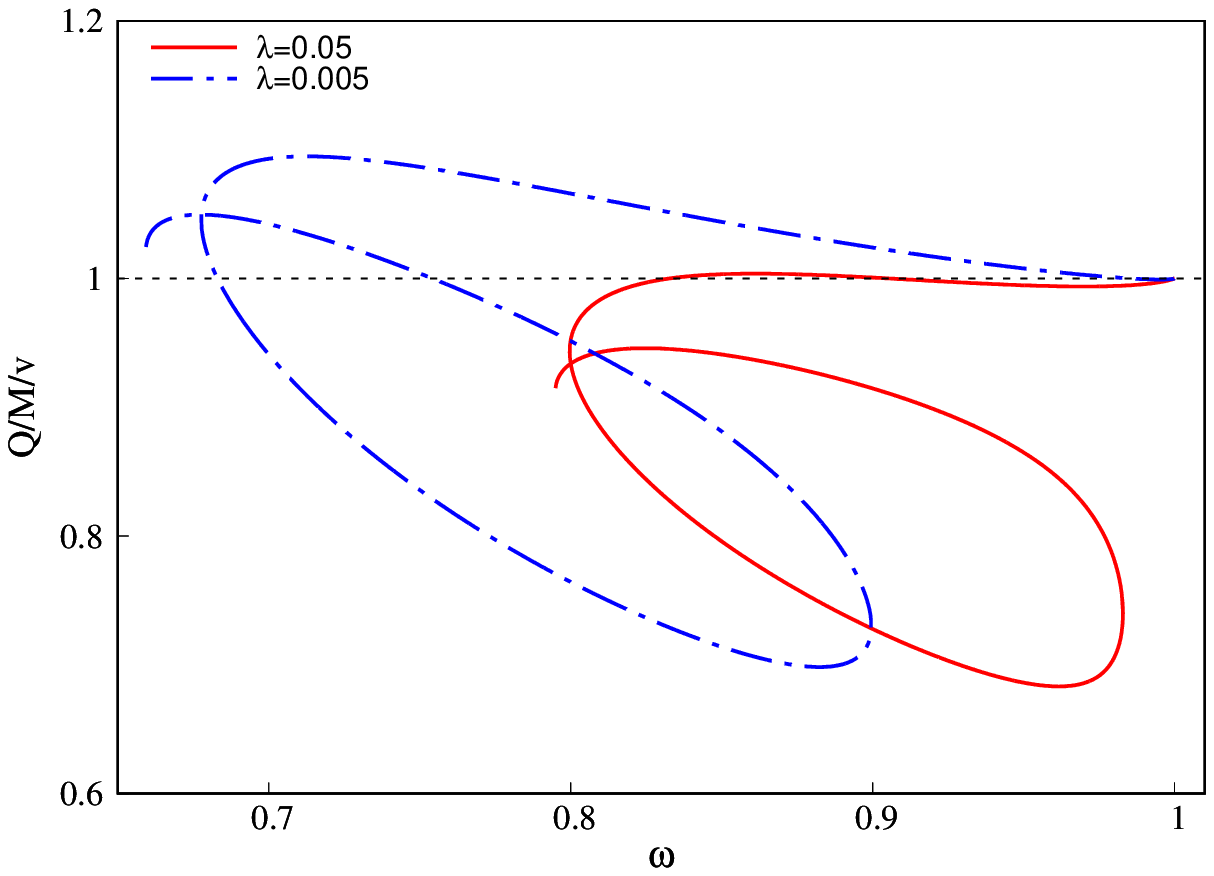}
        \caption{Top panels: mass/energy (red solid line) and Noether charge (blue dashed line) of the Proca-Higgs balls as a function of their frequency $\omega$ for two illustrative values of $\lambda$. Bottom left panel: a comparison between the mass/energy for the two values of the coupling $\lambda=0.005$ (red solid line) and $\lambda=0.05$ (blue dashed line). The self-intersection of the mass/energy curve is a feature observed for all computed $\lambda$.  Bottom right panel: the ratio $Q/(M/v)$ - when it is smaller than one, the Proca-Higgs balls should be unstable against fission. }
        \label{fig:massfreqcharge}
\end{figure}

For the Proca-Higgs balls, the only tunable parameter in the model is $\lambda$.  Figure \ref{fig:massfreqcharge}  exhibits the domain of existence of the fundamental solutions\footnote{All solutions presented in this work correspond to the fundamental solutions (or ground state), for which the vector potential temporal component $f(r)$ has the minimum number of nodes (one). Solutions with higher number of nodes exist, corresponding to excited states, but we shall not report them here.}  for two illustrative values of $\lambda$, in mass $vs.$ frequency diagrams. Several features are salient. Firstly, as $\omega\rightarrow 1$ the mass/energy tends to diverge. This is a well known feature of $Q$-ball models. Secondly, and unlike $Q$-balls, as one scans the domain of solutions one finds a self-intersection of the mass/energy curve (and also of the Noether charge one). Thirdly, unlike $Q$-balls, for which a minimum frequency emerges wherein the mass/energy again diverges, so far  we were not able to compute a minimal frequency for Proca-Higgs balls. In Figure \ref{fig:massfreqcharge} we have plotted the solutions curve only until the third branch, for better visualization. Finally, the criterion for energetic instability $Q<M/v$ is obeyed in a larger part of the parameter space as $\lambda$ increases.

\begin{figure}[h!]
	\centering
	\includegraphics[width=0.47\textwidth]{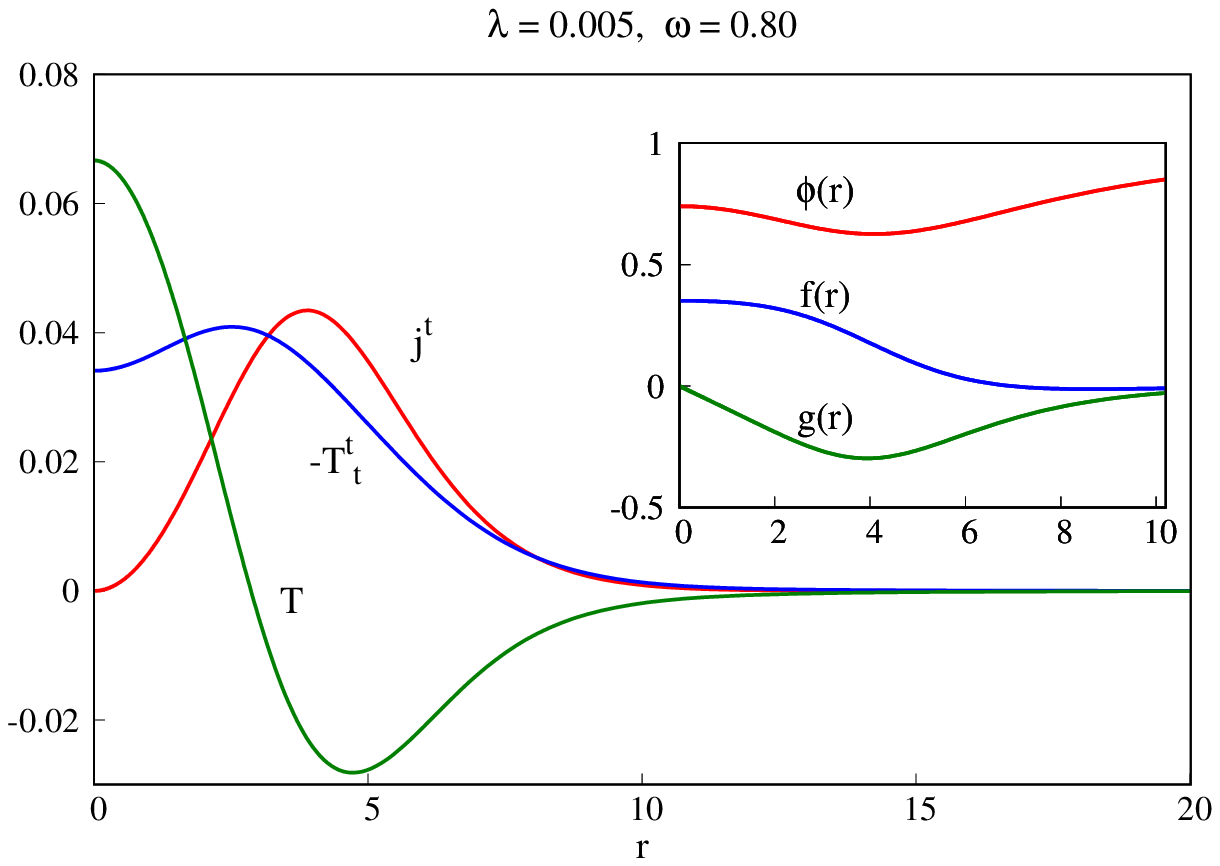}
 \centering
	\includegraphics[width=0.47\textwidth]{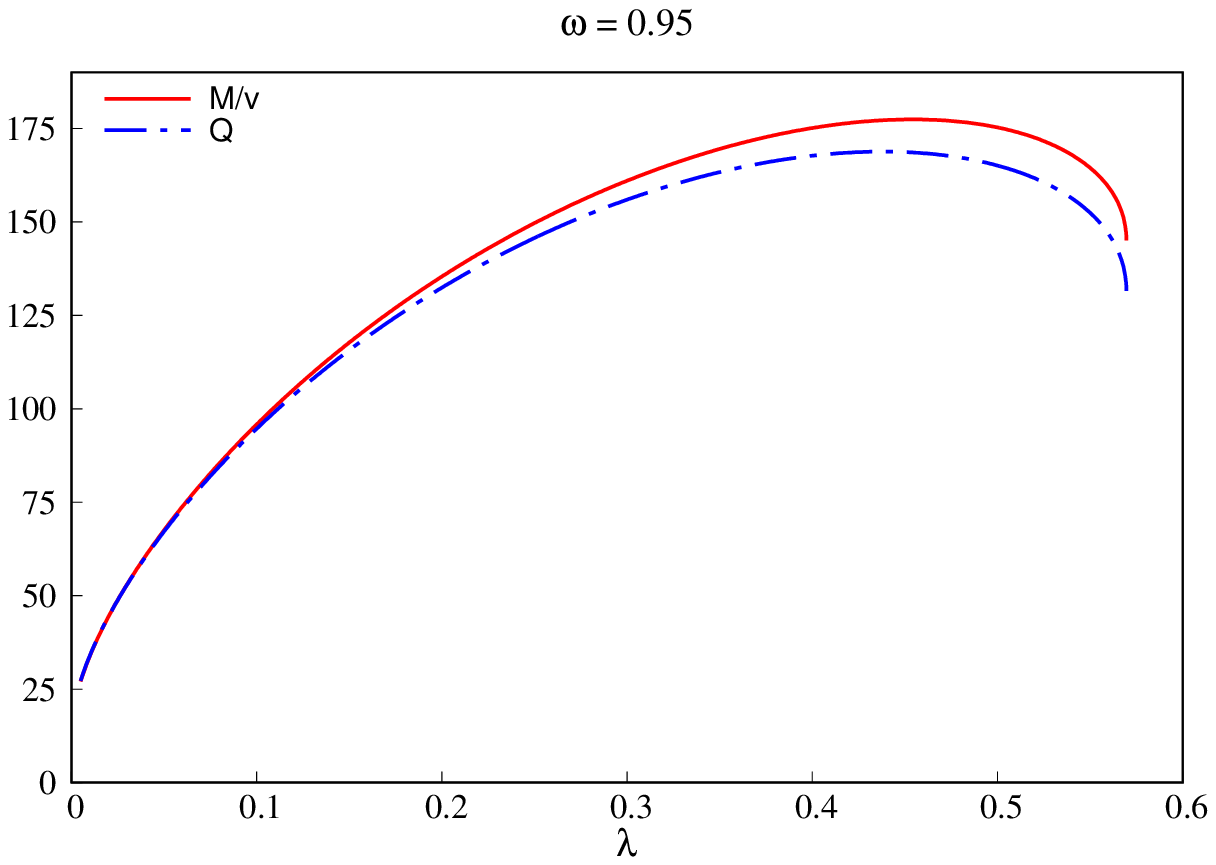}
	\caption{ (Left panel) Radial profiles of the mass/energy and Noether charge densities, energy-momentum trace and (in the inset) vector and scalar functions, for the Proca-Higgs ball with $\omega=0.80$ and $\lambda=0.005$.  The solution lies on the first branch and has $M = 48.26$,  $Q = 51.45$. For this particular solution, the maximum value of $f$, $f=0.351$, occurs at the origin, while its minimum, $f=-0.0126$, occurs at $r=8.756$. (Right panel)  Mass/energy (solid red line) and Noether charge (blue dashed line) as functions of $\lambda$ for $\omega=0.95$.}
	\label{fig:fw8lam005}
\end{figure}

 Differently from $Q$-balls and scalar boson stars, $\phi_0$ is not a good solution label for Proca-Higgs stars. A good parameter is $f_0$, as for Proca stars,  since it grows monotonically along the solutions curve, starting from the maximal frequency limit; thus  it is in a one to one correspondence with the solutions. In this respect, we remark that for a generic solution, $f_0$ is not the maximum of $f(r)$; rather, the maximum occurs in the neighbourhood of the origin. To illustrate the profile functions of a concrete solution we exhibite them for an illustrative Proca-Higgs ball with $\omega=0.80$ and $\lambda=0.005$ in Figure~\ref{fig:fw8lam005} (left panel).

We have performed thorough studies of the solution space for particular values of $\lambda$, including $\lambda=0.05$ and $\lambda=0.005$, scanning for the frequency $\omega$. To illustrate the dependence of the solution space on the coupling constant, we have fixed the frequency, say $\omega=0.95$, and varied $\lambda$ - Figure \ref{fig:fw8lam005} (right panel). Interestingly, the model seems to have an upper bound on the parameter $\lambda$. As discussed in the previous section, increasing $\lambda$ the model tends to the standard (free) Proca model. As shown in \cite{Herdeiro:2016tmi}, the latter does not admit flat spacetime solutions for the considered ansatz. In agreement with this, our numerical results suggest that the solutions cease to exist above some particular value of $\lambda$, which depends on the frequency $\omega$.

\section{Proca-Higgs stars}
\label{sec4}
\subsection{The equations and asymptotic behaviors}
\label{sec4-eq}

Let us now turn on gravity $\alpha\neq 0$ and study the Proca-Higgs stars of the model~\eqref{action}. Again we focus on spherical symmetry and thus the full ansatz is given by~\eqref{matteransatz} and \eqref{metric}. The scaled matter field equations~\eqref{v2}-\eqref{s2} then yield:
\begin{equation}
\dfrac{d}{dr}\left\{r^2\left[f'-\omega g\right]e^{F_1-F_0}\right\}=r^2 e^{3F_1-F_0}f\phi^2\,, \qquad \omega g-f'=\dfrac{e^{2F_0}g\phi^2}{\omega}\,,
\end{equation}
\begin{equation}
\phi''= \left(g^2-e^{2 F_1-2 F_0}f^2 -\lambda  e^{2 F_1}\right)\phi-\left(\dfrac{2}{r}+ F_0'+F_1'\right)\phi' +\lambda  e^{2 F_1} \phi^3\,.
\end{equation}
The Lorenz-like gauge condition for these equations becomes
\begin{equation}
g'+\omega f e^{2 F_1-2 F_0}=- g \left(F_0'+F_1'+\frac{2 \phi'}{\phi}+\frac{2}{r}\right) \ .
\end{equation}
The scaled Einstein equations~\eqref{s2}, are combined as in the following: the (t,t) component and $(r,r)+(\theta,\theta)-\dfrac{(t,t)}{2}$; leading into two equations to be solved
\begin{multline}\label{einstein1}
F_1''=- \dfrac{F_1'^2}{2}-\frac{2 F_1'}{r}+\frac{\alpha ^2}{2}  \left\{e^{-2 F_{0}} \left(\left(f'-w g\right)^2+f^2 e^{2 F_{1}} \phi^2+e^{2 F_{0}} g^2 \phi^2\right)-\dfrac{\lambda}{2}  e^{2 F_{1}}\left(\phi^2-1\right)^2- \phi'^2\right\} \ ,
\end{multline}
\begin{multline}\label{einstein0}
F_0''= -2 F_{0}' F_{1}'-F_{0}'^2-\frac{3 F_{0}'}{r}-\frac{1}{2} F_{1}'^2-\frac{F_{1}'}{r}+\\+\frac{\alpha ^2}{2}  \left\{e^{-2 F_{0}} \left(\left(f'-w g\right)^2+5 f^2 e^{2 F_{1}} \phi^2+e^{2 F_{0}} g^2 \phi^2\right)+ \phi'^2-\dfrac{3}{2} \lambda  e^{2 F_{1}} \left(\phi^2-1\right)^2\right\} \ .
\end{multline}
The Noether charge and mass are now
\begin{align}
&Q=4\pi\int_{0}^{\infty}\frac{e^{F_0+F_1} g^2\phi^2}{\omega} r^2dr\,,\\ &M=2\pi v\int_{0}^{\infty}e^{F_0+ F_1}\left[(f'-\omega g)^2+\phi^2\left(f^2e^{2 F_1} +g^2 e^{2 F_0} \right)+\dfrac{e^{2F_0+2F_1}\lambda}{2}(\phi^2-1)^2+e^{2 F_0} \phi'^2\right]r^2dr\, ,
\end{align}
whereas the virial identity is
\begin{multline}
\int_{0}^{\infty} \Bigg\{-\dfrac{e^{-2F_1}}{2}\phi'^2-\dfrac{3\lambda}{4}\left(\phi^2-1\right)^2+\dfrac{1}{2}\phi^2\left[3 e^{-2F_0}f^2 + e^{-2F_1}g^2\left(-1+e^{2F_0}\dfrac{\phi^2}{\omega^2}\right)\right]+\\
\frac{e^{-2 F_1} F_1' \left(2 F_0'+F_1'(r)\right)}{2 \alpha^2}\Bigg\}e^{F_0+3F_1}r^2dr=0\,,
\end{multline}

The solutions reported in this Section obey the virial identity up to errors
of order $10^{-5} - 10^{-6}$, where, as in the case of balls, the larger errors occur inside of the spiral.

Again, the numerical integration requires an analysis of the asymptotic behaviour of the relevant functions.  Close to the origin, $r\approx0$, the different profile functions read
\begin{align}
&F_0(r)=F_{00}+\frac{\alpha ^2}{12}  e^{2 F_{10}-2 F_{00}} \left[4 f_0^2 \phi_0^2-e^{2 F_{00}} \lambda  \left(\phi_0^2-1\right)^2\right]r^2+\mathcal{O}(r^3)\,,\\
&F_1(r)=F_{10}-\frac{\alpha ^2}{24}  e^{2 F_{10}-2 F_{00}} \left[2 f_0^2 \phi_0^2+e^{2 F_{00}} \lambda  \left(\phi_0^2-1\right)^2\right]r^2+\mathcal{O}(r^3)\,,\\
&\phi(r)=\phi_0-\frac{\phi_0}{6}  e^{2 F_{10}-2 F_{00}} \left[f_0^2-e^{2 F_{00}} \lambda  \left(\phi_0^2-1\right)\right]r^2+\mathcal{O}(r^3)\,,\\
&f(r)=f_0+\frac{f_0}{6}  e^{2 F_{10}-2 F_{00}} \left(e^{2 F_{00}} \phi_0^2-\omega^2\right)r^2+\mathcal{O}(r^3)\,,\\
&g(r)=-\frac{f_0 \omega}{3}e^{2F_{10}-2F_{00}}r+\mathcal{O}(r^3)\,.
\end{align}

Moreover, since we want to describe asymptotically flat solutions, the matter field behavior for large $r$ is still described by equations \eqref{phiinf},\eqref{finf} and \eqref{ginf}, while the behavior of the metric functions is

\begin{equation}
e^{2F_0(r)}=1- \dfrac{2MGv}{r}+\cdots,\qquad e^{-2F_1(r)}=1- \dfrac{2MGv}{r}+\cdots,
\end{equation}
where the parameter $M$ can be identified as the ADM mass.

\subsection{Numerical Results}\label{Asymptotic_grav}

Let us now present the numerical results for the spherical Proca-Higgs stars. We use the same numerical framework as in the case of balls. The solutions presented here have typical errors of order $10^{-5}-10^{-6}$.

\begin{figure}[h!]
	\centering
	\includegraphics[width=0.47\textwidth]{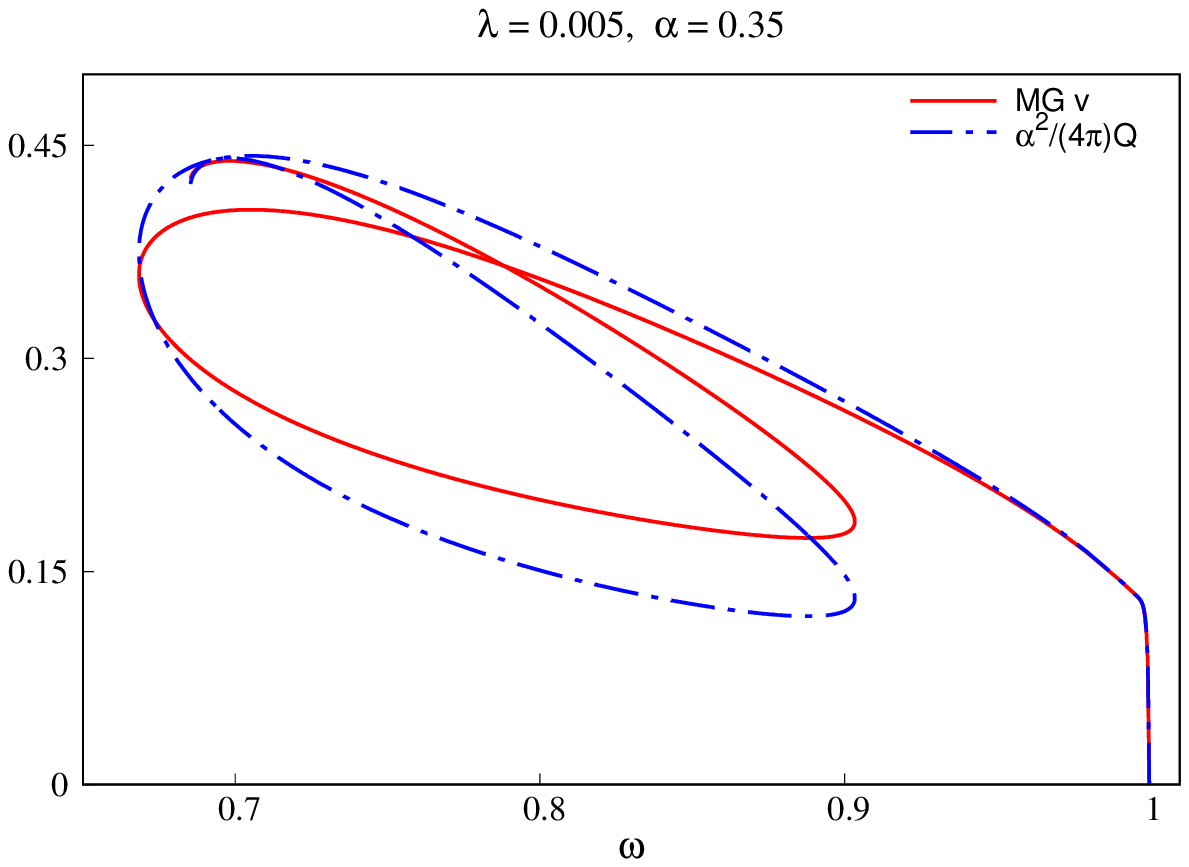}
 \centering
	\includegraphics[width=0.47\textwidth]{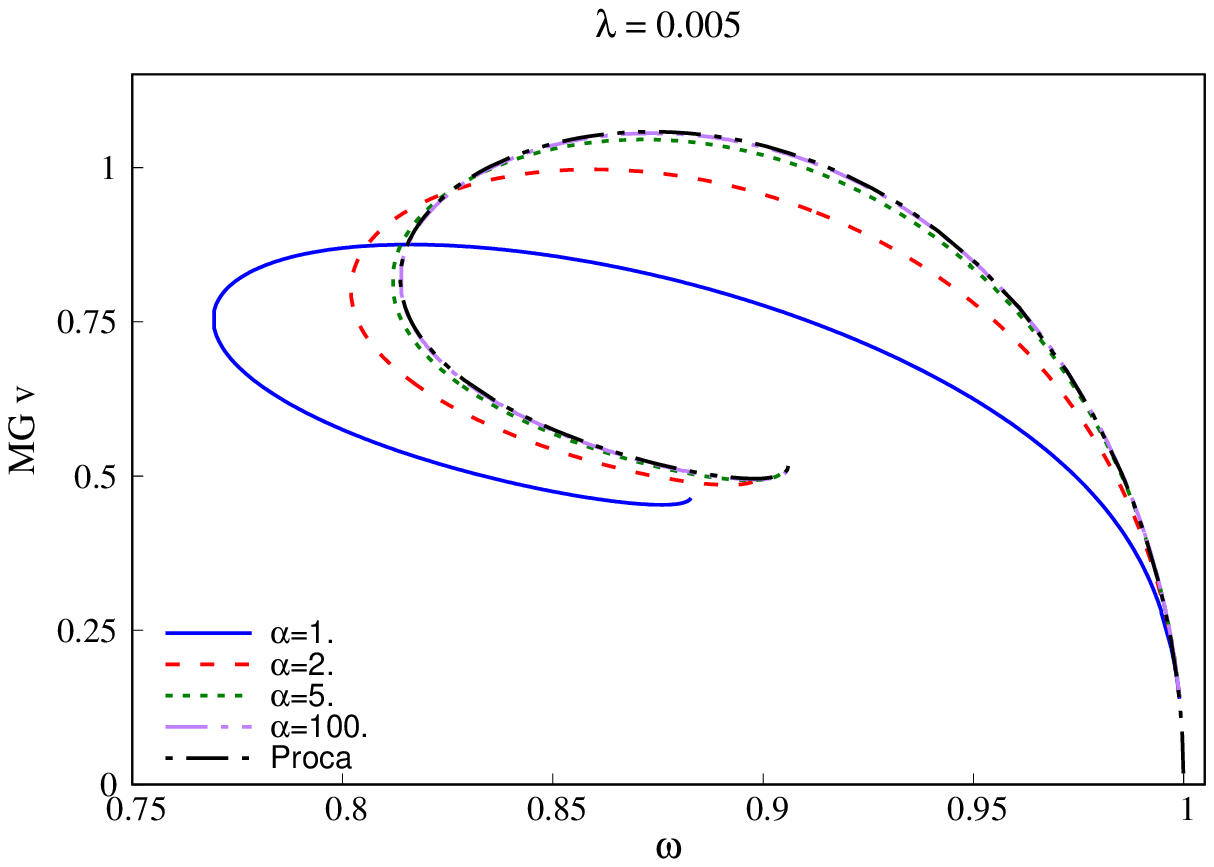}
	\caption{(Left panel) Mass (red solid line) and Noether charge (blue dashed line)  $vs.$ $\omega$, for Proca-Higgs stars with $\lambda=0.005$ and $\alpha=0.35$. (Right panel) Mass $vs.$ $\omega$ curve for Proca-Higgs stars with  $\lambda=0.005$ and four different values of $\alpha$ as well as for the mini-Proca stars.}
	\label{fig:masschargegravfreq_lamb005}
\end{figure}

To scan the parameter space of the Proca-Higgs stars, we must sweep $(\lambda,\alpha)$ and $\omega$. Let us start by fixing $\lambda$ and increasing $\alpha$ - Figure~\ref{fig:masschargegravfreq_lamb005}. The left panel shows a key difference when gravity is turned on - the mass of the solutions is regularised (as compared to balls) as the maximal frequency is reached. In this limit, often called the Newtonian limit, the stars become more dilute and their mass tends to zero. The plot also manifests that the self intersecting mass (and Noether charge) curve remains, even when gravity is turned on, at least for sufficiently small $\alpha$. For sufficiently large $\alpha$, on the other hand, this feature is absent. In fact, for large enough $\alpha$ the Higgs field is almost frozen at its vev, and Proca-Higgs stars  mass curve approaches that of mini-Proca stars - Figure~\ref{fig:masschargegravfreq_lamb005} (right panel).

\begin{figure}[h!]
     \centering
     \begin{subfigure}[h]{0.47\textwidth}
         \centering
         \includegraphics[width=\textwidth]{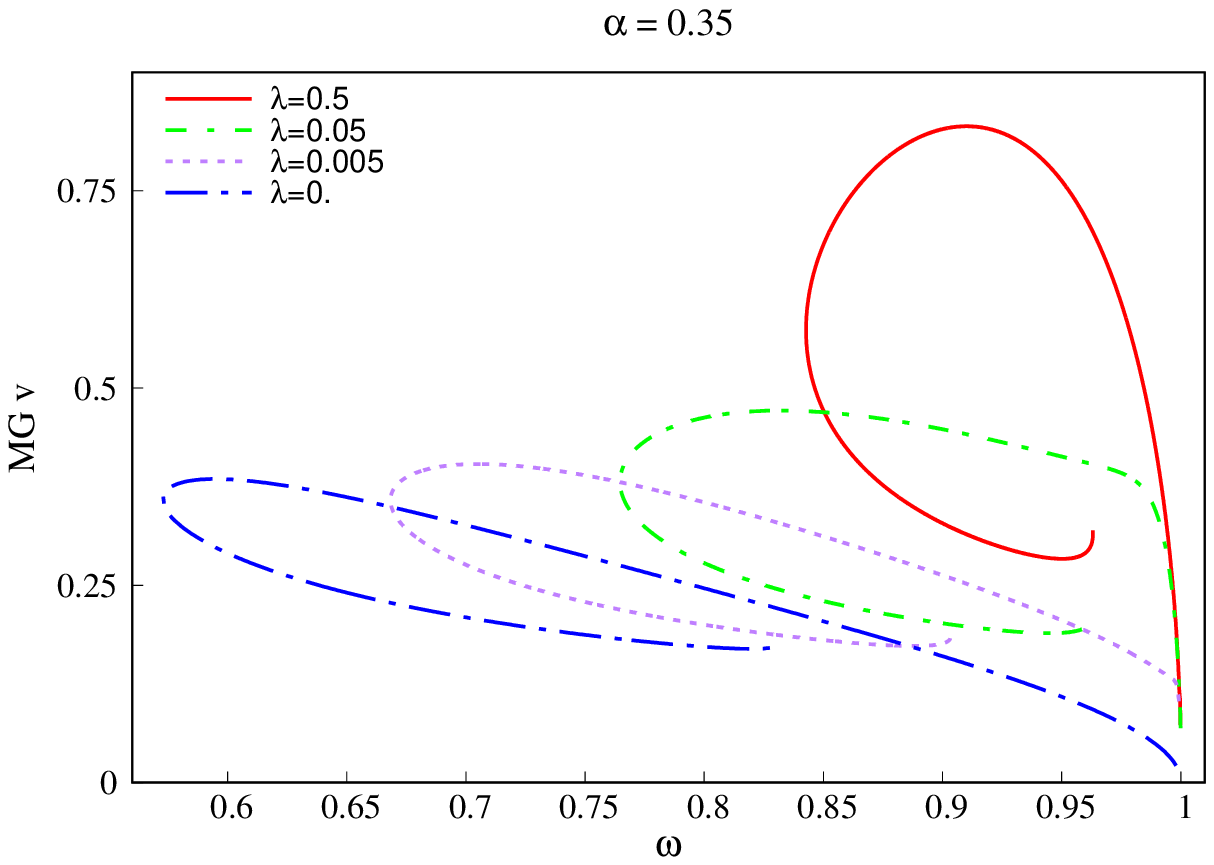}
         \caption{}
         \label{fig:massfreqcharge0005grav}
     \end{subfigure}
     \hfill
     \begin{subfigure}[h]{0.47\textwidth}
         \centering
         \includegraphics[width=\textwidth]{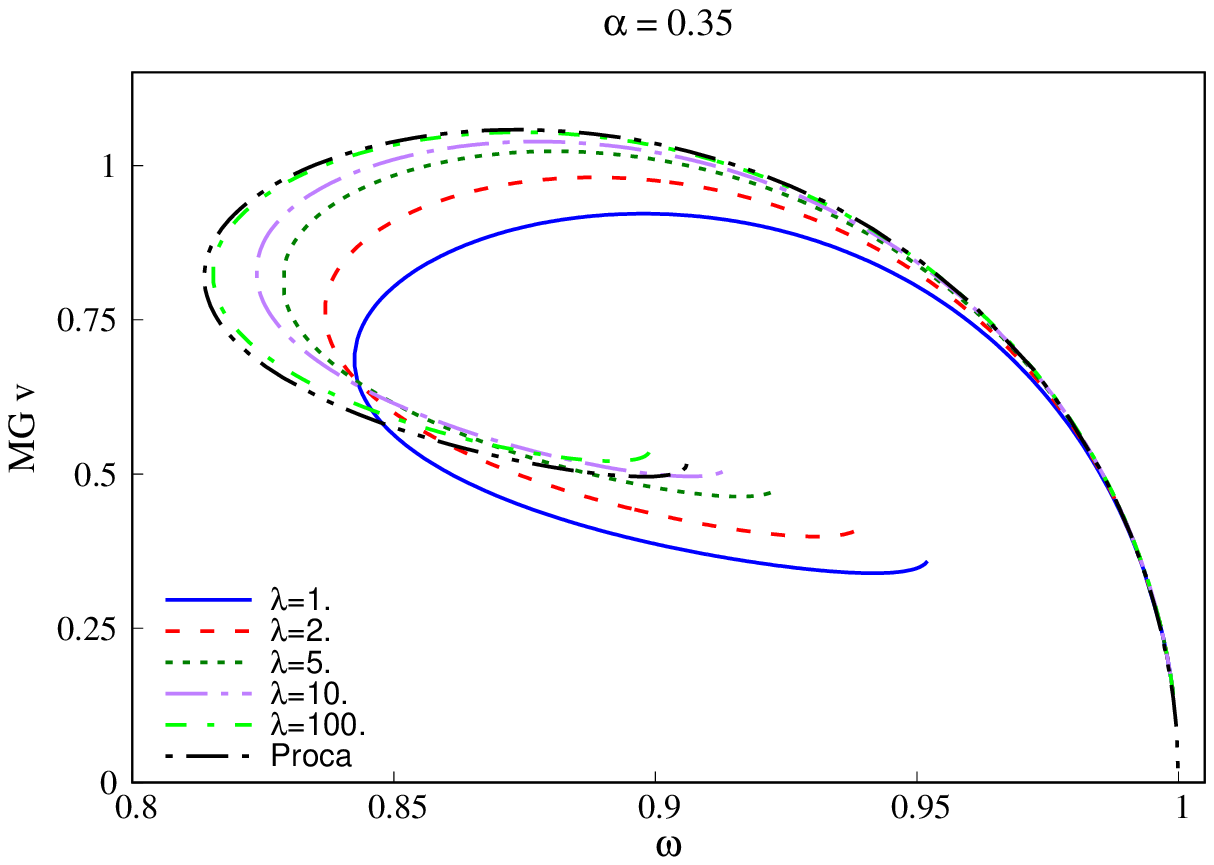}
         \caption{}
         \label{fig:massfreqcharge005grav}
     \end{subfigure}
     \begin{subfigure}[h]{0.47\textwidth}
     \centering
	\includegraphics[width=\textwidth]{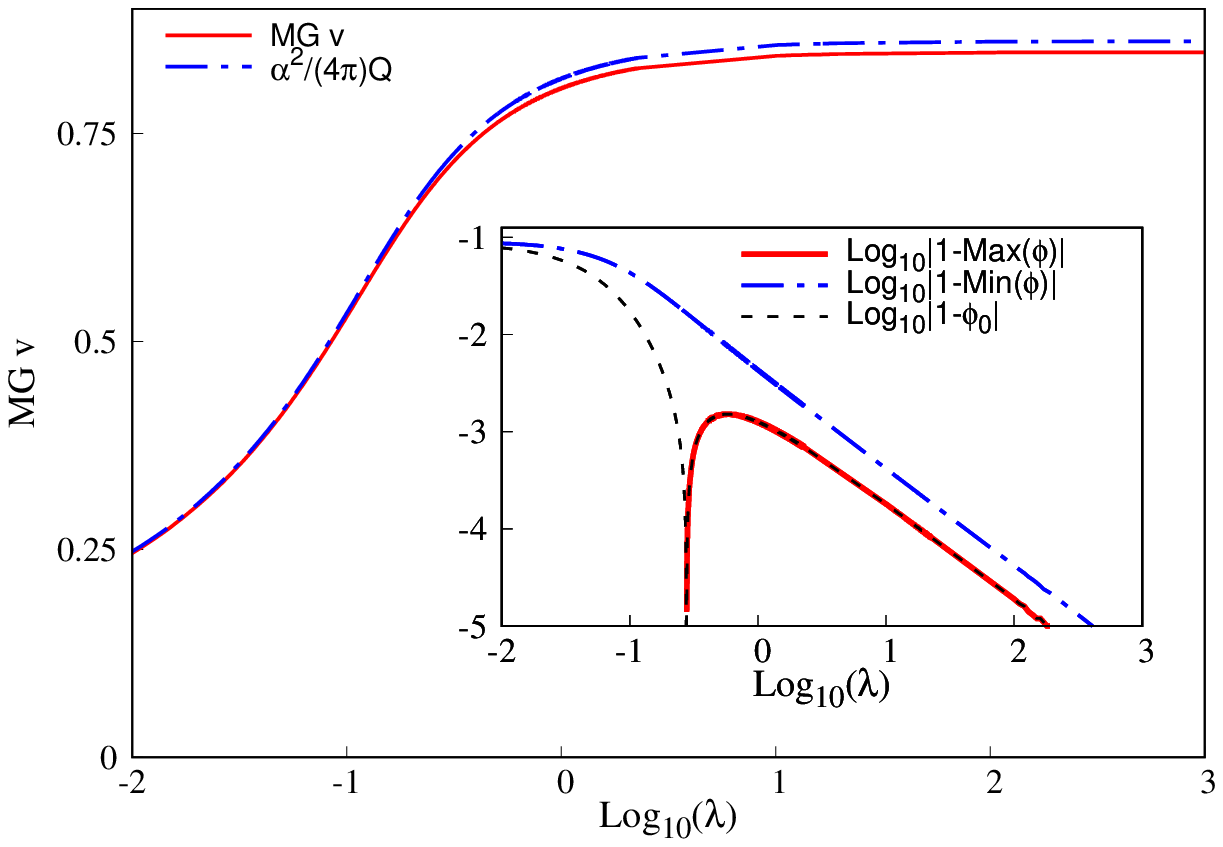}
  \end{subfigure}
   \begin{subfigure}[h]{0.47\textwidth}
  \centering
	\includegraphics[width=\textwidth]{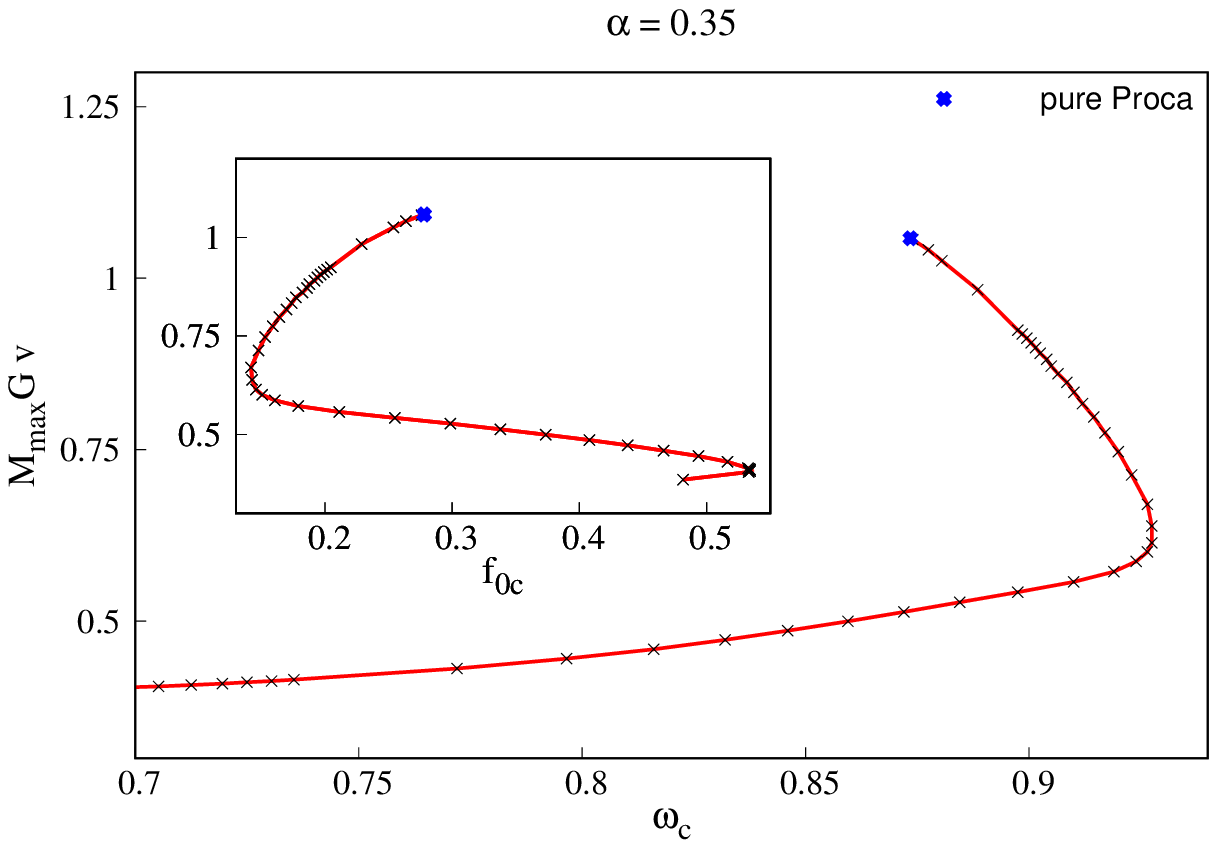}
   \end{subfigure}
        \caption{(Top left and right) Variation of the Proca-Higgs stars Mass $vs.$ $\omega$ curve with $\lambda$, for $\alpha=0.35$. The solutions tend to mini-Proca stars $\lambda$  becomes large. (Bottom left) The mass (red solid line) and Noether charge (blue dashed line) of Proca-Higgs stars solutions, as functions $\lambda$ for $\omega=0.95$ and $\alpha=0.35$. The mass approaches that of mini-Proca stars. The inset shows how much the scalar field deviates from $1$: for large values of $\lambda$, $\phi_0$, Min($\phi$) and Max($\phi$) are all $\simeq 1$. (Bottom right) Variation of the critical frequency, $\omega_{c}$, at which the maximal Mass is attained (for fixed $\alpha$ and $\lambda$)  when $\lambda$ is varied and   $\alpha=0.35$:  black crosses are data points, whilst the red continuous line is an interpolation. For the computed solutions, the maximal mass monotonically increases with $\lambda$, tending to that of mini-Proca stars (blue cross).}
        \label{fig:phaselam}
\end{figure}

Let us now fix $\alpha$ and vary $\lambda$ instead - Figure~\ref{fig:phaselam}. A feature that is lost when gravity is turned on is the apparent maximum of $\lambda$ described in the last section. Solutions  exist for arbitrarily high $\lambda$. This is to be expected, from the reasoning presented before, and the solutions should approach, again, the mini-Proca stars for $\lambda\rightarrow \infty$. In fact, this is what we see in Figure~\ref{fig:phaselam} (top right panel). Again, the Higgs field freezes at its vev in this limit (bottom left panel). Comparing the top left and right panels of Figure~\ref{fig:phaselam}, one notices, however, a change in trend: for low (high) $\lambda$, increasing $\lambda$, increases (decreases) the minimum frequency at which the first branch of solutions ends, and the first backbending of the solution space occurs. The same non-monotonic behaviour with $\lambda$ also occurs for the critical frequency at which the maximal mass occurs (bottom right panel).

\begin{figure}[h!]
	\centering
	\includegraphics[width=0.47\textwidth]{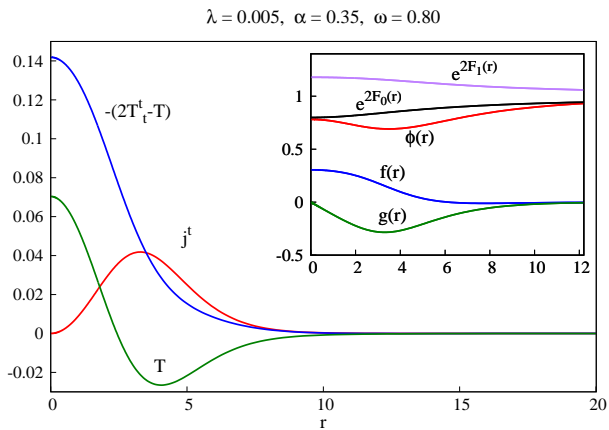}
	\caption{Radial profiles of the mass and Noether charge densities, energy-momentum trace and (in the inset), metric, vector and scalar functions, for the Proca-Higgs stars with $\omega=0.80$,  $\lambda=0.005$ and $\alpha=0.35$.  The solution lies on the first branch and has $M = 36.52$,  $Q = 38.86$. For this particular solution, the maximum value of $f$, $f=0.0305$, occurs at the origin, while its minimum, $f=-0.0109$, occur at $r=7.695$.}
	\label{fig:gravF1w8lam005}
\end{figure}

Finally, we exhibit the profile of the metric and  matter fields profiles, together with some physical quantities, for an illustrative solution, with $\omega=0.80$, $\lambda=0.005$ and $\alpha=0.35$, in Figure~\ref{fig:gravF1w8lam005}.

\section{Compactness and some special geodesics}
\label{sec5}

To get some further insight on the Proca-Higgs stars we have presented in the previous section, we shall now look at their compactness and some special circular geodesics. 

\subsection{Compactness}
As we have seen, Proca-Higgs stars approach mini-Proca stars for both large $\lambda$ and large $\alpha$. Mini-Proca stars are stable from the maximal frequency up to the frequency of the maximal mass~\cite{Brito:2015pxa}. This is a typical property of spherical bosonic stars~\cite{Cunha:2017wao}. Thus we shall call the corresponding part of the first branch of Proca-Higgs stars the ``stable branch", albeit a rigorous stability analysis is beyond the scope of this paper.

Both scalar boson stars and Proca stars become more compact along the stable branch, from the Newtonian limit up to the maximal mass. Following the literature (see $e.g$~\cite{Herdeiro_2015}), we define this compactness in terms of the effective (areal) radius $R_{99}$, which contains $99\%$ of the total mass of the star. Note that these consideration are done in terms of the areal radius $R$, which has a geometric meaning, and that connects to the isotropic radius $r$ as $R_{99}=e^{2F_1}r_{99}$. Then, the inverse compactness is defined by
\begin{equation}
    \text{Compactness}^{-1}=\frac{R_{99}}{2M_{99}},
\end{equation}
where $M_{99}=0.99M$.

\begin{figure}[h!]
     \centering
     \begin{subfigure}[h]{0.47\textwidth}
         \centering
         \includegraphics[width=\textwidth]{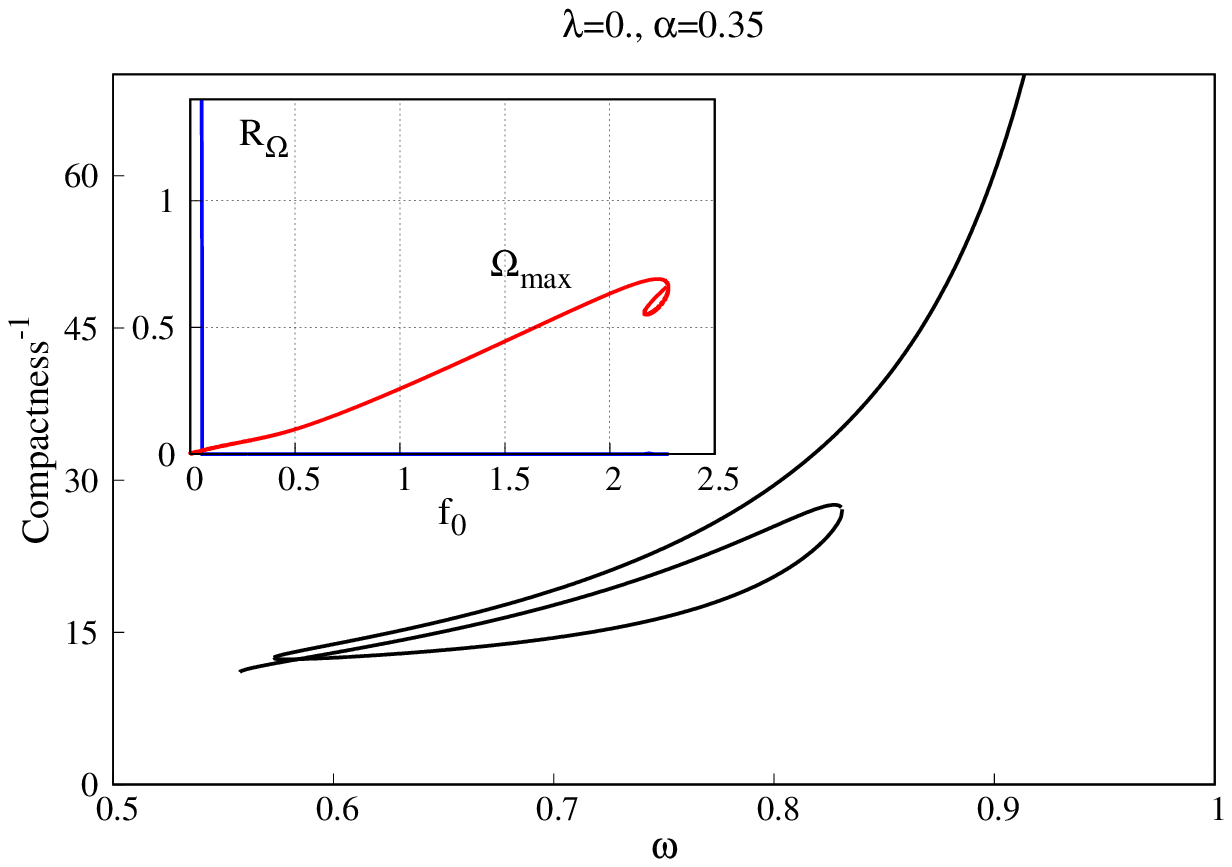}
         \caption{}
         \label{fig:compact_lam0}
     \end{subfigure}
     \hfill
     \begin{subfigure}[h]{0.47\textwidth}
         \centering
         \includegraphics[width=\textwidth]{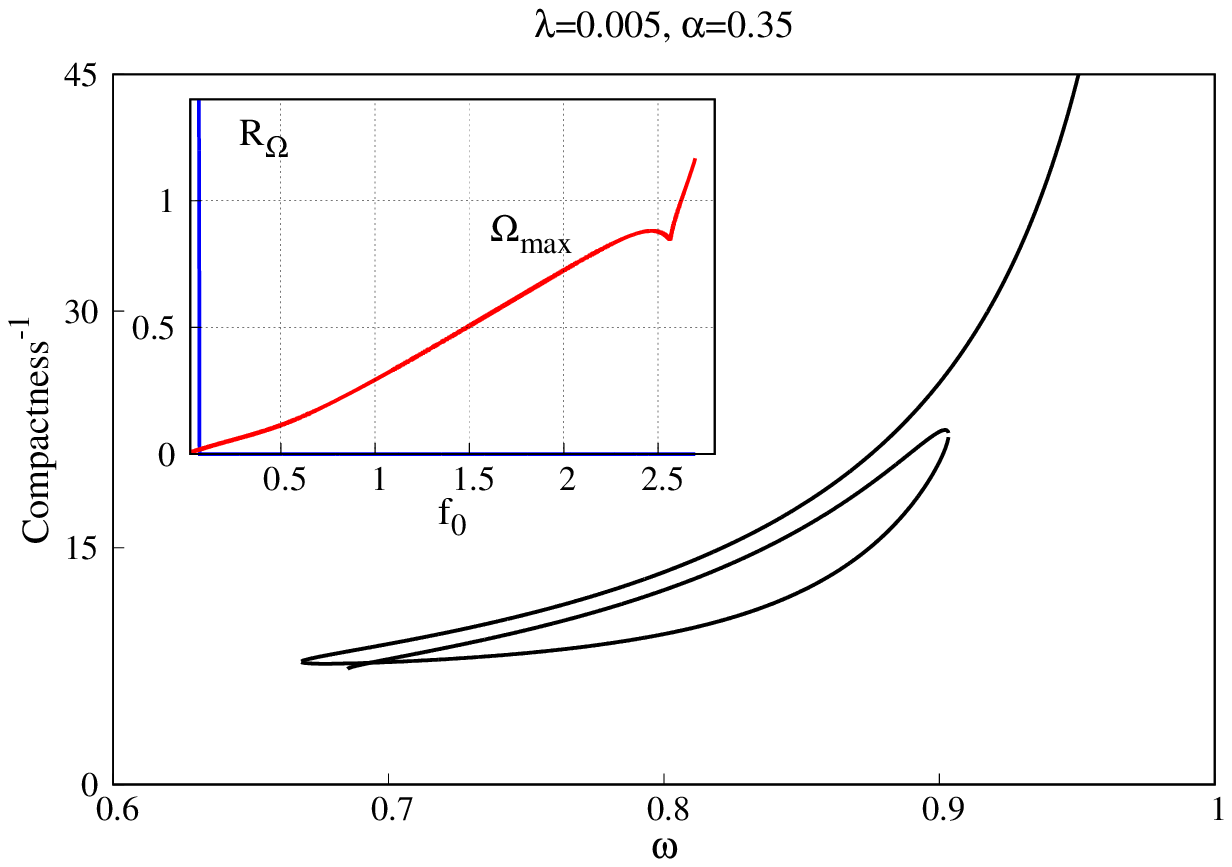}
         \caption{}
         \label{fig:compact_lam0005}
     \end{subfigure}
    \begin{subfigure}[h]{0.47\textwidth}
         \centering
         \includegraphics[width=\textwidth]{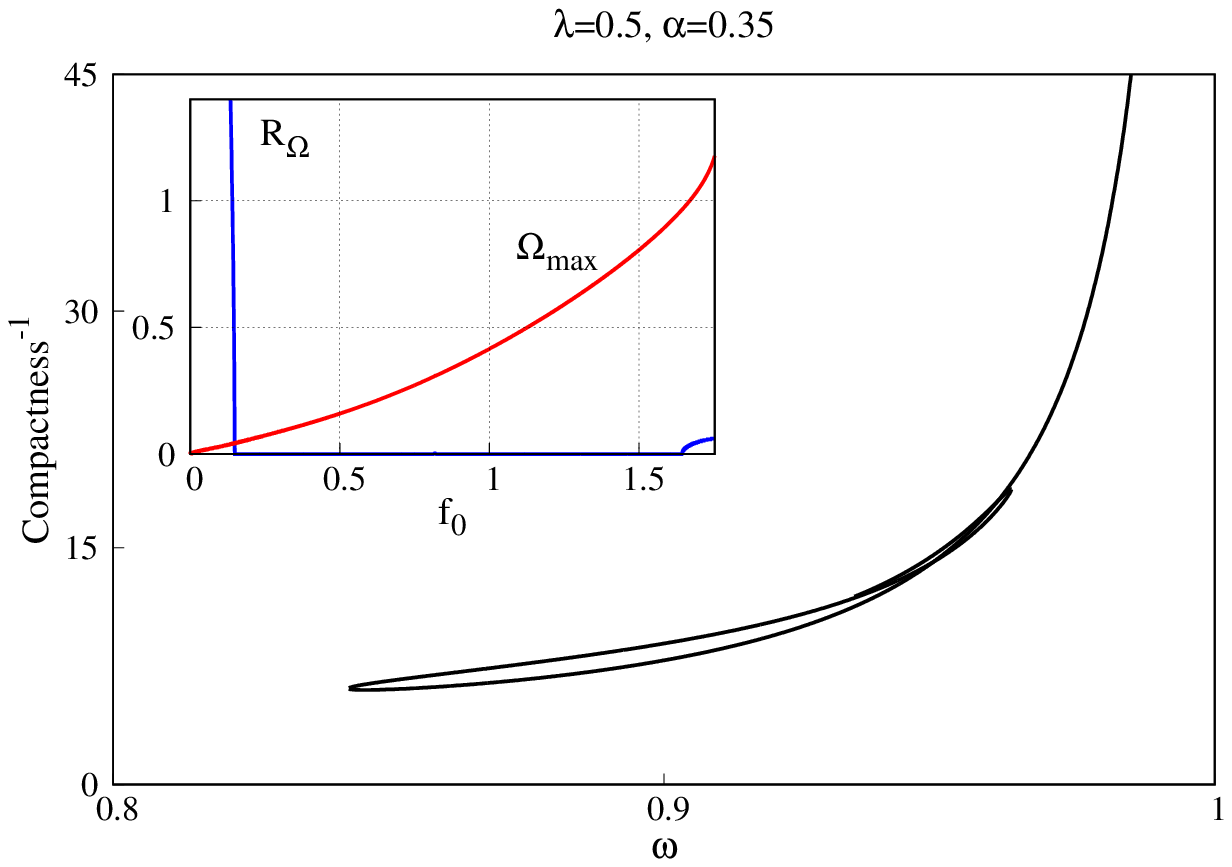}
         \caption{}
         \label{fig:compact_lam05}
     \end{subfigure}
     \hfill
     \begin{subfigure}[h]{0.47\textwidth}
         \centering
         \includegraphics[width=\textwidth]{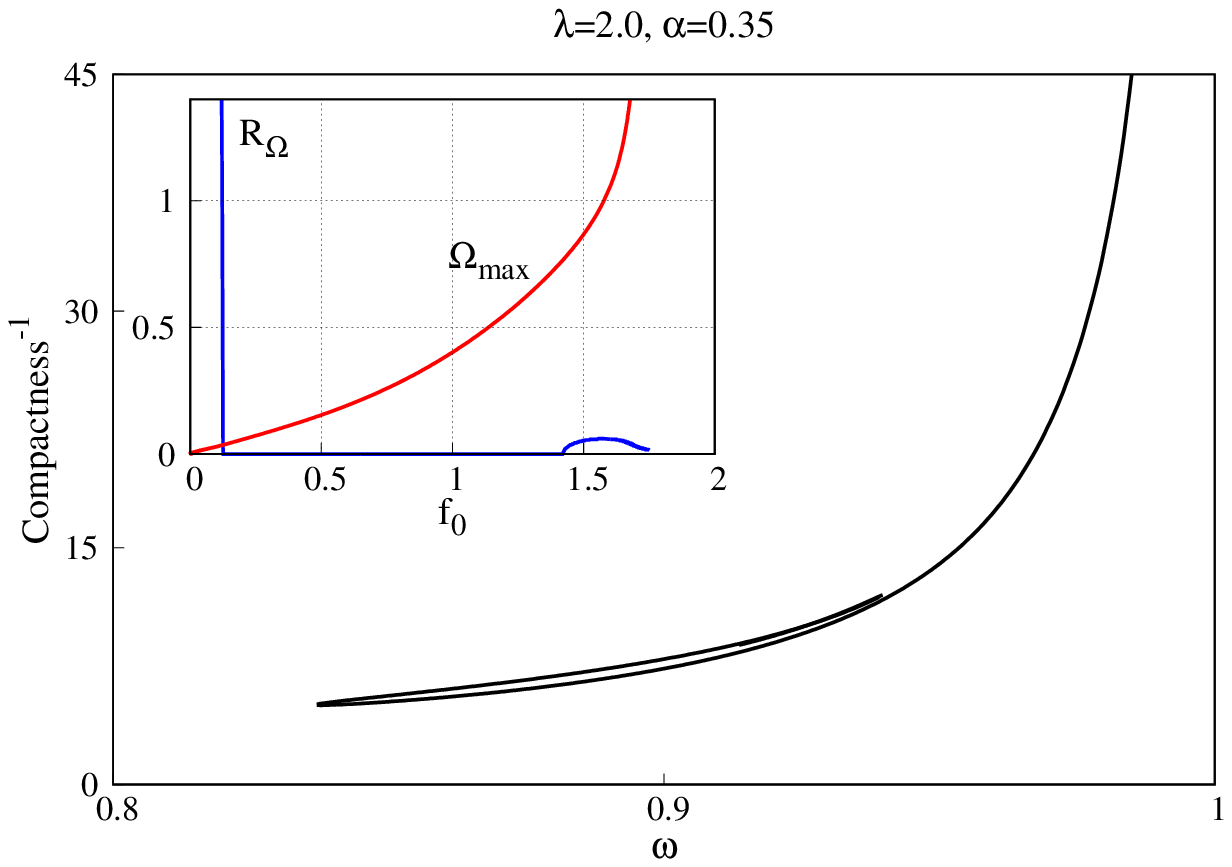}
         \caption{}
         \label{fig:compact_lam2}
     \end{subfigure}
        \caption{ The inverse compactness as a function of the angular frequency $\omega$. In the inset, we plot the areal radius of the maximal angular velocity along TCOs, $R_\Omega$ (blue solid line) and the corresponding value of
the angular velocity $\Omega_{\text{max}}$ (red solid line), as a function of the parameter $f_0$.}
        \label{fig:compact}
\end{figure}

The Higgs-Proca case is no different from the aforementioned models: moving along the whole first branch the stars become more compact - Figure~\ref{fig:compact} (main panels). Along secondary branches, the compactness may increase or decrease, but the stars therein become fairly compact, albeit never as compact as a Schwarzschild black hole.

\begin{figure}[h!]
         \includegraphics[width=0.49\textwidth]{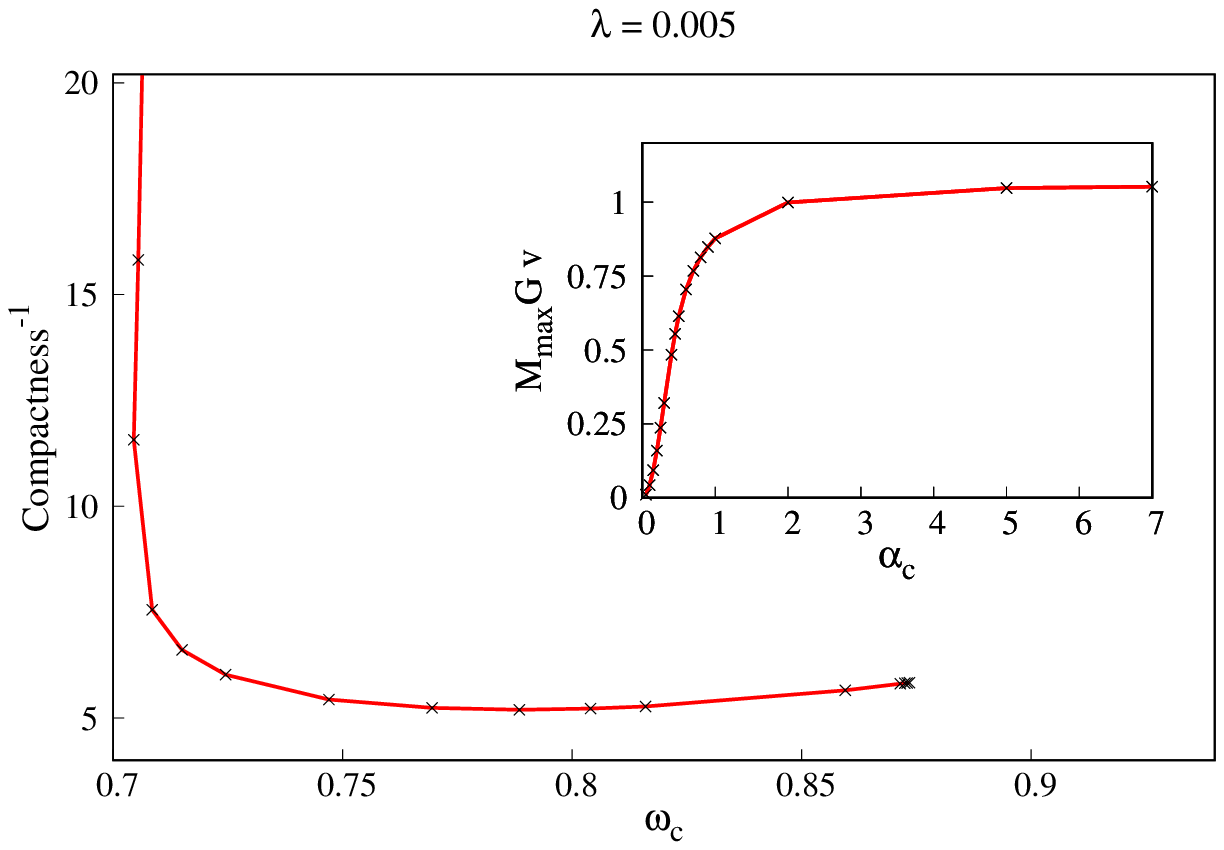}
         \includegraphics[width=0.49\textwidth]{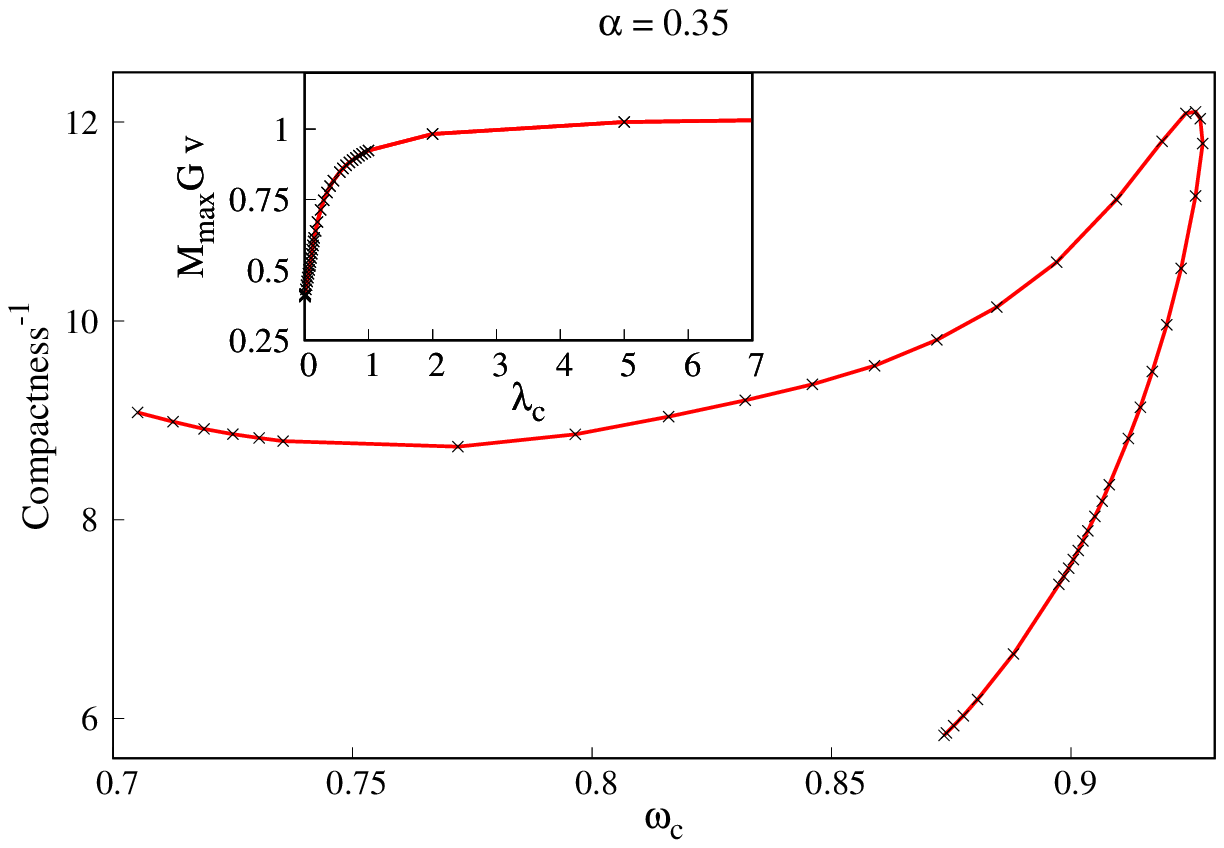}

        \caption{Inverse compactness of the maximal mass solution along the parameter space. Left panel: $\lambda$ is fixed  and $\alpha$ is increased along the red solid line from left to right. Right panel: $\alpha$ is fixed and $\lambda$ increased along the red solid line from left to right until the back-bending and then from top to bottom. The insets show the corresponding mass of the maximal mass solutions.}
        \label{fig:compact2}
\end{figure}

One can also compare the compactness of the maximal mass solution as one moves along the $(\lambda,\alpha)$ parameter space. This is illustrated in Figure~\ref{fig:compact2}, fixing an illustrative value of $\lambda$ (left panel) or $\alpha$ (right panel). One observes that the compactness of the maximal mass solution is not monotonic. Fixing $\lambda$, on the one hand, the maximal mass solution initially increases considerably in compactness when increasing $\alpha$, reaching a maximal value, and then decreasing slightly, tending to the value of the mini-Proca case. Fixing $\alpha$, on the other hand, the compactness of the maximal mass solution initially increases slightly when increasing $\lambda$, reaching a maximal value, and then decreasing  towards a global minimum of compacness in solution space (for the maximal mass solution). Then it increases tending to the value of the mini-Proca case. It is worth noticing that along both sequences of solutions, the maximal mass always increases, as shown in the inset of both panels in Figure~\ref{fig:compact2}. 

\subsection{Circular geodesics}

Let us now turn to the study of circular geodesics. Special spacetime circular geodesics include light rings (LRs) and innermost stable circular orbits (ISCOs). The existence of these features in a horizonless star-like compact object could make them black holes foils, see $e.g.$~\cite{Herdeiro:2021lwl}, which justifies their interest. More recently, it was argued that even without an ISCO, a horizonless spacetime could imitate the shadow of a black hole, under some conditions, if its timelike circular orbits (TCOs) reached a maximal angular velocity at some non-zero radius~\cite{Olivares:2018abq}. This rationale was explored in~\cite{Herdeiro:2021lwl} to argue that some mini-Proca stars in the stable branch could actually imitate the shadow of a Schwarzschild black hole. Since the Proca-Higgs stars connect to mini-Proca stars, we expect the same feature to hold for the former. The following analysis will corroborate this hypothesis.

We consider null/time-like circular geodesics on the geometry~\eqref{metric}, which are planar. The radial geodesic equation, on the equatorial plane, for a massive (massless) particle, is
\begin{equation}
  V_k(r) \equiv  \dot{r}^2=k\, e^{-2F_1}+e^{-2F_0-2F_1} E^2-\dfrac{e^{-4F_1}l^2}{r^2} \ ,
\end{equation}
where $E$, $l$ represent the particle’s energy and angular momentum and the dot represents the derivative with respect to an affine parameter;  $k= -1, 0$ for time-like or null geodesics, respectively. Then,  a particle on a  circular orbit at $r = r^{cir}$ simultaneously satisfies the conditions (see \cite{Delgado:2021jxd} for a more detailed discussion): 
\begin{subequations}
\begin{align}
    V_k(r^{cir})=0 \ ,\label{Vorbit}\\
    V_k'(r^{cir})=0 \ .\label{vprimeorbit}
\end{align}
\end{subequations}
The stability of such orbits is dictated by the second derivatives
\begin{equation}
    V''_k(r^{cir}) >0 \Leftrightarrow \text{unstable};\qquad V''_k(r^{cir}) <0 \Leftrightarrow \text{stable}.
\end{equation}

Let us first consider the time-like case. For TCOs, the angular velocity of the particle along the geodesic is given by:
\begin{equation}
    \Omega=\dfrac{d\,\varphi}{d\, t}=\frac{e^{F_0-F_1}\sqrt{F_0'}}{\sqrt{r(1+r F_1')}}\Bigg|_{r^{cir}}.
\end{equation}
Let $R_\Omega$ denote the areal radius which maximizes the angular velocity along TCOs. Following \cite{Herdeiro:2021lwl}, we define

\begin{equation}
    \xi\equiv \frac{R_{\text{ISCO}}}{R_\Omega},
\end{equation}
where $R_\text{ISCO}$ corresponds to the ISCO radius for a  Schwarzschild black hole with the same mass. This measures the ability for a compact object to mimic the accretion flow of a Schwarzschild black hole~\cite{Olivares:2018abq,Herdeiro:2021lwl}.  Again, these consideration are done in terms of the areal radius $R$.
As mentioned before, $f_0$ grows monotonically along the $M$ $vs.$ $\omega$ solutions line. Then, we can introduce the quantity
\begin{equation}
    \chi=\dfrac{f_0}{f_0(M_{\text{max}})},
\end{equation}
as an indicator of how close a given solution, characterized by $f_0$, is from the end of the stable branch.

We now consider null geodesics. For LRs, $k=0$, the combination of the conditions \eqref{Vorbit} and \eqref{vprimeorbit} yields to the algebraic equation
\begin{equation}
    r F_0'(r)-r F_1'(r)-1=0 \ .
\end{equation}
The relations just established can now be computed for the numerical solutions of Proca-Higgs stars.

Our analysis of the numerical data revealed no LRs or ISCO are present for the studied solutions in the stable branch. This is similar to the mini-Proca stars case, where such features appear only for more compact (and unstable) solutions~\cite{Cunha:2017wao}.  On the other hand, as for mini-Proca stars, a maximum of the angular velocity for TCOs emerges at a non vanishing radius within solutions in the stable branch. In particular, we may look for $R_\Omega=6M$, so that  $\xi=1$. Such ``special" solutions would in principle mimic the shadow of a Schwarzschild black hole, since the accretion flow stalls, leaving an empty inner hole, that under some observation conditions would be essentially degenerate with the Schwarzschild shadow~\cite{Herdeiro:2021lwl}. In the following table we provide the numerical data for such special solutions for different values of $\lambda$ and  $\alpha$.

\bigskip

\begin{center}
\centering
\begin{tabular}{|cccccccc|}
\hline
\multicolumn{8}{|c|}{``Special" solution along the parameter space}                           \\ \hline
\multicolumn{1}{|c|}{$\alpha$} & \multicolumn{1}{c|}{$\lambda$} & \multicolumn{1}{c|}{$ \omega$} & \multicolumn{1}{c|}{Mass}  & \multicolumn{1}{c|}{$\chi(\xi=1)$} & \multicolumn{1}{c|}{$r_{\Omega}$} & \multicolumn{1}{c|}{$\Omega$} & Compactness$^{-1}$ \\ \hline
\multicolumn{1}{|c|}{0.35}     & \multicolumn{1}{c|}{2.}        & \multicolumn{1}{c|}{0.9369}    & \multicolumn{1}{c|}{0.890} & \multicolumn{1}{c|}{0.362}         & \multicolumn{1}{c|}{4.797}        & \multicolumn{1}{c|}{0.025}    & 11.146           \\
\multicolumn{1}{|c|}{0.35}     & \multicolumn{1}{c|}{0.5}       & \multicolumn{1}{c|}{0.9386}    & \multicolumn{1}{c|}{0.800} & \multicolumn{1}{c|}{0.503}         & \multicolumn{1}{c|}{4.333}        & \multicolumn{1}{c|}{0.026}    & 11.938           \\
\multicolumn{1}{|c|}{0.35}     & \multicolumn{1}{c|}{0.005}     & \multicolumn{1}{c|}{0.9414}     & \multicolumn{1}{c|}{0.217} & \multicolumn{1}{c|}{0.125}         & \multicolumn{1}{c|}{1.273}        & \multicolumn{1}{c|}{0.017}    & 39.468           \\
\multicolumn{1}{|c|}{0.35}     & \multicolumn{1}{c|}{0}        & \multicolumn{1}{c|}{0.8914}     & \multicolumn{1}{c|}{0.168} & \multicolumn{1}{c|}{0.115}         & \multicolumn{1}{c|}{0.970}        & \multicolumn{1}{c|}{0.013}    & 55.436     \\
 \multicolumn{1}{|c|}{0.50} & \multicolumn{1}{c|}{0.005} & \multicolumn{1}{c|}{0.9407} & \multicolumn{1}{c|}{0.359} & \multicolumn{1}{c|}{0.121}  & \multicolumn{1}{c|}{2.053}  & \multicolumn{1}{c|}{0.021} & 24.568\\
  \multicolumn{1}{|c|}{2.0} & \multicolumn{1}{c|}{0.005} & \multicolumn{1}{c|}{0.9367} & \multicolumn{1}{c|}{0.847} & \multicolumn{1}{c|}{0.257}  & \multicolumn{1}{c|}{4.556}  & \multicolumn{1}{c|}{0.025} & 11.692\\
    \multicolumn{1}{|c|}{1.0} & \multicolumn{1}{c|}{1.0} & \multicolumn{1}{c|}{0.9362} & \multicolumn{1}{c|}{0.916} & \multicolumn{1}{c|}{0.313}  & \multicolumn{1}{c|}{4.882}  & \multicolumn{1}{c|}{0.025} & 11.001\\
    \hline
        \multicolumn{1}{|c|}{Proca} & \multicolumn{1}{c|}{-} & \multicolumn{1}{c|}{0.936} & \multicolumn{1}{c|}{0.925} & \multicolumn{1}{c|}{0.248}  & \multicolumn{1}{c|}{4.970}  & \multicolumn{1}{c|}{0.025} & 10.939\\
 \hline
\end{tabular}
\end{center}

\bigskip

Let us close this section by mentioning that we have found some numerical evidence of solutions with LRs, but for the parameters explored on the 4th branch of solutions.

\section {Spherical hairy black holes?}
\label{sec6}
It is possible to prove that spherical scalar boson stars cannot be put in equilibrium with a black hole horizon to generate a ``hairy" black hole~\cite{Pena:1997cy}. The same can be shown for spherical Proca stars~\cite{Herdeiro:2016tmi}. One may wonder if the Proca-Higgs stars can be different. Here, we follow the methodology in \cite{Herdeiro:2016tmi} to construct a modified P\~{e}na-Sudarsky theorem~\cite{Pena:1997cy} and establish that Proca-Higgs stars cannot be put in equilibrium with a black hole horizon. 

We consider a spherically symmetric line element in Schwarzschild-like coordinates (unlike the isotropic coordinates used before~\eqref{metric}), and with parameterization
\begin{equation}
d s^{2}=-\sigma^{2}(r) N(r) d t^{2}+\frac{d r^{2}}{N(r)}+r^{2} d \Omega_{2} \ , \qquad N(r) \equiv 1-\frac{2 m(r)}{r} \ .
\end{equation}
The ansatz we consider for the complex Proca potential and real scalar is written as before, $cf.$~\eqref{matteransatz}. The Proca field equations  yield
\begin{align}
\frac{d}{d r}\left\{\frac{r^{2}\left[f^{\prime}(r)-\omega g(r)\right]}{\sigma(r)}\right\}=\frac{\phi(r)^{2} r^{2} f(r)}{\sigma(r) N(r)} \ , \\
f^{\prime}(r)=\omega g(r)\left(1-\frac{\phi(r)^{2} \sigma^{2}(r) N(r)}{\omega^{2}}\right) \ .
\label{fequ}
\end{align}

The Lorenz-like condition determines $f (r)$ in terms of the other
functions:
\begin{equation}
f(r)=-\frac{\sigma(r) N(r)}{\omega r^{2} \phi(r)^2} \frac{d}{d r}\left[r^{2}\phi(r)^2 \sigma(r) N(r) g(r)\right] \ ,
\end{equation}
or 
\begin{equation}
\frac{d}{d r}\left[r^{2}\phi(r)^2 \sigma(r) N(r) g(r)\right]=-\frac{\omega r^{2} f(r)\phi(r)^2}{\sigma(r) N(r)} \ .
\end{equation}

The $T_t\,^t$ component of the energy-momentum tensor – the energy density – reads
\begin{multline}
-T_t\,^t=	-\frac{\omega g(r) f'(r)}{\sigma (r)^2}+\frac{f'(r)^2}{2 \sigma (r)^2}+\frac{f(r)^2 \phi(r)^2}{2 N(r) \sigma (r)^2}+\\\frac{\omega^2 g(r)^2}{2 \sigma (r)^2}+\frac{1}{2} g(r)^2 N(r) \phi(r)^2+\frac{\lambda }{4}+\frac{1}{2} N(r) \phi'(r)^2+\frac{1}{4} \lambda  \phi(r)^4-\frac{1}{2} \lambda  \phi(r)^2 \ .
\label{Ttt}
\end{multline}

To establish the no-Proca-Higgs hair theorem for spherical black holes, let us assume the existence of a regular black hole solution of the above equations. Then, the geometry would possess a non-extremal horizon at, say, $r = r_H > 0$, which requires that
\begin{equation}
N(r_H)=0 \ .
\end{equation}
The regularity of the horizon implies that the energy density of the Proca-Higgs field is finite there. From~\eqref{Ttt}  one can see that this implies
\begin{equation}
f(r_H)=0 \qquad \text{or}\qquad \phi(r_H)=0 \ .
\end{equation}
Moreover, neither the functions might diverge for $r=r_H$. Let us first consider the case in which $	f(r_H)=0$. Then, the function $f(r)$ starts from zero at the horizon and remains strictly positive (or negative) for some $r$-interval. Defining 
\begin{equation}
P(r)\equiv 1-\frac{\phi(r)^{2} \sigma^{2}(r) N(r)}{\omega^{2}} \ ,
\end{equation}
we realize from~\eqref{fequ} that the sign of $f'$ depends on the sign of $P$ and $g$. Moreover, $P(r_H)=1$ and for large $r$, $P$ becomes negative, since $N\rightarrow 1$, $\sigma\rightarrow 1$ and $\phi\rightarrow 1$, but we also have $\omega<1$ to ensure an exponential decay of the Proca field at infinity. Then, let $r_1>r_H$ be the first zero of $P$ after the horizon. Hence, in the horizon vicinity, the sign of $f'$ equals that of $g$, which thus determines the sign of $f$. Then, let the first zero of $g$ be at $r_2$, hence, $g$ is either positive or negative in the interval $r_H\le r\le r_2$. Consequently, $f'$ and $f$ are strictly positive or strictly negative in the interval $r_H\le r \le r*\equiv\text{min}\left\{r_1,r_2\right\}$. Now, integrating  the gauge equation for any $r$ in the interval $r_H\le r \le r*$, we get:

\begin{equation}
r^{2}\phi(r)^2 \sigma(r) N(r) g(r)=-\omega\int_{r_H}^{r}\frac{ x^{2} f(x)\phi(x)^2}{\sigma(x) N(x)}d\,x
\end{equation}

The equation above introduces a contradiction. If $f(r)$ is negative, then the R.H.S. is positive and $g$ must be positive. But we have seen that $g$ and $f$ must have the same sign in the interval $r_H\le r \le r*$. Hence, the only solution possible is $g=f=0$. 

Instead, if we assume $\phi(r_H)=0$, then, by means of the scalar field equation we see that all derivatives of $\phi$ must be zero at $r=r_H$. Hence, $\phi$ would not be an analytical function at this point.

This establishes there are no spherically symmetric black holes with Proca-Higgs hair.

\section{Conclusions}
\label{sec7}
In this paper we have introduced a Proca-Higgs model, as a vector version of the Friedberg-Lee-Sirlin model, that allows a complex vector field to acquire mass dynamically. The Proca-Higgs model can be seen as a UV completion of a self-interacting Proca model, free of hyperbolicity issues, that reduces to the free Proca model in some limits, albeit the latter is never a consistent truncation of the Proca-Higgs model.

We have constructed solitonic solutions of the Proca-Higgs model, both as non-gravitating solitons on flat spacetime (Proca-Higgs balls) and as self-gravitating solitons that curve spacetime (Proca-Higgs stars). The existence of Proca-Higgs balls is a differentiating feature, as compared to the standard (free) Proca model. One may interpret it as a result of the effective self-interactions introduced by the scalar-vector coupling. These flat spacetime solitons cease to exist, however, for sufficiently strong self-interactions, that mandate the scalar field to be essentially frozen at its vev.

The space of solutions of Proca-Higgs balls has some qualitative differences with that, say, of $Q$-balls, notably the fact that different solutions with the same mass and frequency (and number of radial nodes of the appropriate functions) can exist. This remains true for Proca-Higgs stars, with sufficiently low couplings. For sufficiently strong gravitational coupling or self-interactions, the Proca-Higgs stars tend to mini-Proca stars. The latter provide the upper limits of mass an Nother charge for the Proca-Higgs stars.

We have also shown that the physical properties of the Proca-Higgs stars bifurcate from those of mini-Proca stars (again, in the limit of large couplings), for instance, in terms of the compactness or of the structure of special circular geodesics. Finally, we have shown that no spherical black hole horizon can be in equilibrium with Proca-Higgs stars, as for spherical Proca and scalar boson stars.

The Proca-Higgs model and solitons introduced here allow for many applications and generalizations. Let us just mention two that may be worth pursuing. First, dynamical studies of the solitons are of interest, both to assess single-soliton stability and multi-soliton dynamics. In the gravitating case, the possibility of extracting gravitational waves from collisions of such objects is an interesting one, possibly for phenomenological studies to compare with real data. Second, rotating solitons (both balls and stars) should exist and can be computed. In particular, the stars should again bifurcate, for large couplings, from spinning mini-Proca stars, which are dynamically robust~\cite{Sanchis-Gual:2019ljs}, making the corresponding spinning Proca-Higgs solitons potentially interesting. Moreover, in the spinning case, the no-hair theorem for spherical black holes with bosonic star hair is circumvented, and, as for scalar~\cite{Herdeiro:2014goa,Herdeiro:2015gia} and vector~\cite{Herdeiro:2016tmi,Santos:2020pmh}, spinning bosonic stars, black hole generalizations should be possible.

\section*{Acknowledgement}
 E.S.C.F. would like to thanks Alexandre Pombo and Jorge Delgado for their support. We would also like to thank N. Santos for discussions. This work is supported  by the  Center for Research and Development in Mathematics and Applications (CIDMA) through the Portuguese Foundation for Science and Technology (FCT -- Fundac\~ao para a Ci\^encia e a Tecnologia), references  UIDB/04106/2020 and UIDP/04106/2020.  
The authors acknowledge support  from the projects CERN/FIS-PAR/0027/2019, PTDC/FIS-AST/3041/2020,  CERN/FIS-PAR/0024/2021 and 2022.04560.PTDC.  
This work has further been supported by  the  European  Union's  Horizon  2020  research  and  innovation  (RISE) programme H2020-MSCA-RISE-2017 Grant No.~FunFiCO-777740 and by the European Horizon Europe staff exchange (SE) programme HORIZON-MSCA-2021-SE-01 Grant No.~NewFunFiCO-101086251. E.S.C.F. is supported by the FCT grant PRT/BD/153349/2021 under the IDPASC Doctoral Program. Computations have been performed at the Argus and Blafis cluster at the U. Aveiro.

\appendix

\section{First law}

In this Appendix we construct a first law-like relation for the soliton solutions of model~\eqref{action}. To do so, let us assume the existence of an everywhere time-like Killing vector. Then, consider the Komar mass \cite{Bardeen:1973gs}
\begin{equation}
M=-\int_{\Sigma}\left(2 T_{\nu}\,^{\mu}-T \delta_{\nu}^{\mu}\right) K^{\nu} d \Sigma_{\mu} \ ,
\end{equation}
where $K^{a}$ is a time-like Killing vector. Let us first focus on the term
\begin{equation}\label{Tint}
\int_{\Sigma} T_{\nu}\,^{\mu} K^{\nu} d \Sigma_{\mu}=\int_{\Sigma}\left(\frac{1}{2}
( \mathcal{F}^{\mu\gamma }\bar{\mathcal{F}}_{\nu \gamma}
+\bar{\mathcal{F}}^{\mu \gamma} \mathcal{F}_{\nu \gamma}
)+\phi^2\frac{1}{2}
(
\mathcal{A}_{\nu}\bar{\mathcal{A}}^{\mu}
+\bar{\mathcal{A}}_{\nu}\mathcal{A}^{\mu}
)\right)K^{\nu} d \Sigma_{\mu}+ L_{m} \ ,
\end{equation}
where we have defined the Lagrangian of the matter field
\begin{equation}
    L_{m}=\int_{\Sigma}\mathcal{L}_{m}K^{\mu} d \Sigma_{\mu}\,, \qquad\ .
\end{equation}
\begin{equation}
\mathcal{L}_{m}=-\frac{1}{4}\mathcal{F}_{\alpha\beta}\mathcal{\bar{F}}^{\alpha\beta}
-\frac{1}{2}\phi^2\mathcal{A}_\alpha\bar{\mathcal{A}}^\alpha
-\frac{1}{2} \partial_\alpha \phi \partial^\alpha \phi
-U(\phi) \ .
\end{equation}
Now, we can write the Lie derivative of $\mathcal{A}_\mu$ as 
\begin{equation}
    \mathcal{L}_{K}(\mathcal{A}_\mu)=K^{\nu}F_{\nu\mu}+\nabla_{\mu}(K^\nu\mathcal{A}_\nu ) \ .
\end{equation}

Differently from what is traditionally done in the literature on the thermodynamics of stationary gravitating objects, here the electromagnetism potentials depend on time. This gives the calculation an extra step (see \cite{Compere:2006my} for a good review). Hence, by considering a harmonic time dependence for the vector potential, we also have
\begin{equation}
    \mathcal{L}_{K}(\mathcal{A}_\mu)=-i \omega \mathcal{A}_\mu \ .
\end{equation}
Thus, by neglecting boundary terms and assuming that the equations of motion are satisfied, Eq. \eqref{Tint} can be written as

\begin{equation}
    \int_{\Sigma} T_{\nu}\,^{\mu} K^{\nu} d \Sigma_{\mu}-L_{m}=\dfrac{i}{2}\omega\int_{\Sigma}\left(\mathcal{F}^{\mu \gamma}\bar{\mathcal{A}}_{\gamma}-\bar{\mathcal{F}}^{\mu \gamma}\mathcal{A}_{\gamma}\right)  d \Sigma_{\mu}=-\omega Q \ .
\end{equation}
Thus, the mass reads
\begin{equation}
    M=2 \omega Q-2L_{m}+\int_{\Sigma}T  K^{\mu} d \Sigma_{\mu} \ ,
\end{equation}
Moreover, using the Einstein equation \eqref{Einstein-eqs}, we see that $R=-8\pi G T$. Thus, using that the total lagrangian density of the theory is $L=\dfrac{R}{16\pi G}+\mathcal{L}_{m}$, we can finally write the first law-like relation

\begin{equation}\label{first_int}
    M=2 \omega Q-2 L\ ,
\end{equation}

Now, considering an on-shell variation of the above equation
\begin{equation}
    \delta M=2 \delta(\omega Q)+\delta L \ .
\end{equation}

Here, one should be careful by performing the on-shell variation of $L$ or, more specifically, the on-shell variation of $R$. One should take into account that such term possess second order derivatives of the metric. Hence, the boundary term on infinity should be taken into account. Moreover, one can see that such boundary term is equal to $-\delta M$ \cite{Bardeen:1973gs}. Therefore, the differential mass formula becomes
\begin{equation}\label{first_l}
\delta M=\omega \delta Q \ ,
\end{equation}
as advertised. These equations are still valid in the flat spacetime limit.

As a side note, following \cite{Friedberg:1976me}, we can introduce the following Legendre relations
\begin{equation}
    -L= \dfrac{M}{2}- \omega Q\,, \qquad G\equiv \dfrac{M}{2}-\frac{1}{2}\omega Q\,,\qquad I\equiv \dfrac{Q}{\omega} \ .
\end{equation}
Then, we find the ``thermodynamic'' relations \cite{LAINE1998376,Tsumagari:2008bv}
\begin{equation}
    \dfrac{d M}{d Q}\Bigg|_{L}=\omega\,,\qquad  \dfrac{d L}{d \omega}\Bigg|_{M}=Q\,,\qquad \dfrac{d G}{d I}\Bigg|_{L}=\frac{1}{2}\omega^2 \ .
\end{equation}
We should emphasize that this result is valid both for gravitating systems and for the flat limit. In the latter, the metric $g_{\alpha\beta}$ should just be understood as the Minkowski metric. Explicitly, notice also that in the flat spacetime limit, $E=M$, as discussed in the main text. 

To conclude this section, we call out the attention that the variation considered on \eqref{first_l} are those with fixed $\omega$. To understand how the physical quantities change under variations with respect to $\omega$, then take the derivative of \eqref{first_int} with respect to $\omega$

\begin{equation}
        \dfrac{d M}{d \omega}=2 \omega \dfrac{d Q}{d \omega}+2 Q-2 \dfrac{d L}{d \omega}\ .
\end{equation}

One can manipulate the last term in the right hand side, and by means of the equations of motion, one get the definition of the Noether charge plus the derivative of the mass with respect to the frequency. Hence, we have the relation

\begin{equation}
        \dfrac{d M}{d \omega}= \omega \dfrac{d Q}{d \omega}\ .
\end{equation}

\bibliographystyle{hhieeetr}
\bibliography{biblio}

 \end{document}